\documentclass[apj]{emulateapj}
\usepackage{rotating}
\usepackage{multirow}
\renewcommand{\cite}{\citealp}

\journalinfo{The Astrophysical Journal, accepted on the 13th of February 2012}

\shorttitle{On the evolutionary mass of the LMC eclipsing binary Cepheid CEP0227}
\shortauthors{Prada Moroni et al.}

\begin{document}

\title{On the evolutionary and pulsation mass of Classical Cepheids: III.  
the case of the eclipsing binary Cepheid CEP0227 in the Large Magellanic Cloud}

\author{
P. G. Prada Moroni\altaffilmark{1,2},  
M. Gennaro\altaffilmark{3},  
G. Bono\altaffilmark{4,5}, 
G. Pietrzy\'nski\altaffilmark{6,7}, 
W. Gieren\altaffilmark{6}, 
B. Pilecki\altaffilmark{6,7},  
D. Graczyk\altaffilmark{6}, 
I. B. Thompson\altaffilmark{8} 
}   

\altaffiltext{1}{Dipartimento di Fisica, Universit\`a di Pisa, Largo B. Pontecorvo 2, 56127 Pisa, Italy; 
prada@df.unipi.it}
\altaffiltext{2}{INFN--Pisa, via E. Fermi 2, 56127 Pisa, Italy}
\altaffiltext{3}{Max-Planck-Institut f\"{u}r Astronomie, K\"{o}nigstuhl 17, D-69117, Heidelberg, Germany}
\altaffiltext{4}{Dipartimento di Fisica,  Universit\`a di Roma Tor Vergata, 
Via della Ricerca Scientifica 1, 00133 Roma, Italy}
\altaffiltext{5}{INAF--Osservatorio Astronomico di Roma, Via Frascati 33, 00040 Monte Porzio Catone, Italy}
\altaffiltext{6}{Universidad de Concepci\'on, Departamento de Astronomia, Casilla 160-C, Concepci\'on, Chile}
\altaffiltext{7}{Obserwatorium Astronomiczne Uniwersytetu Warszawskiego, Aleje Ujazdowskie 4, 
00-478 Warszawa, Poland}
\altaffiltext{8}{Carnegie Observatories, 813 Santa Barbara Street, Pasadena, CA 91101-1292, USA}
%

\begin{abstract}
We present a new Bayesian approach to constrain the intrinsic parameters (stellar 
mass, age) of the eclipsing binary system --CEP0227-- in the Large Magellanic 
Cloud.  We computed several sets of evolutionary models covering a broad 
range in chemical compositions and in stellar mass. Independent sets of models 
were also constructed either by neglecting or by including a moderate convective 
core overshooting ($\beta_{ov}$=0.2) during central hydrogen burning phases. 
Sets of models were also constructed either by neglecting or by assuming a 
canonical ($\eta$=0.4,0.8) or an enhanced ($\eta$=4) mass loss rate.
The most probable solutions were computed in three different planes:
luminosity--temperature, mass--radius and gravity--temperature. 
By using the Bayes Factor, we found that the most probable solutions were obtained 
in the gravity--temperature plane with a Gaussian mass prior distribution. 
The evolutionary models constructed by assuming a  moderate convective core 
overshooting ($\beta_{ov}$=0.2) and a canonical mass loss rate ($\eta$=0.4) 
give stellar masses for the  primary (Cepheid) --M=4.14$^{+0.04}_{-0.05}$ M$_\odot$-- 
and for the secondary --M=4.15$^{+0.04}_{-0.05}$ M$_\odot$-- 
that agree at the 1\% level with dynamical measurements. 
Moreover, we found ages for the two components and for the 
combined system --t=151$^{+4}_{-3}$ Myr-- that agree at the 5\% level.  
The solutions based on evolutionary models that neglect the mass loss attain 
similar parameters, while those ones based on models that either account 
for an enhanced mass loss or neglect convective core overshooting have 
lower Bayes Factors and larger confidence intervals. The dependence on 
the mass loss rate might be the consequence of the crude approximation 
we use to mimic this phenomenon.   

By using the isochrone of the most probable solution 
and a Gaussian prior on the LMC distance, 
we found a true distance modulus 
--18.53$^{+0.02}_{-0.02}$ mag-- and a reddening value --E(B-V)= 0.142$^{+0.005}_{-0.010}$ mag-- 
that agree quite well with similar estimates in the literature. 
\end{abstract}


\section{Introduction}

Classical Cepheids are very popular in the recent literature, since they play 
a crucial role to address several open astrophysical problems. The reason is threefold. 
{\em i)}-- They are the most popular primary distance indicators, since they are bright 
and can be easily identified. 
{\em ii)}-- They are excellent tracers of intermediate-mass stars, since their 
evolutionary status is well defined (central helium burning, blue loops).    
{\em iii)}-- They are fundamental laboratories to constrain the micro (equation of 
state, opacity, cross sections) and macro (mass loss, mixing, convective transport) 
physics adopted in both evolutionary and pulsation models. 

Dating back to the seminal investigation by \citet[][and references therein]{hof64} 
and by  \citet{bec79}, the evolutionary properties of intermediate-mass have been 
investigated in a series of theoretical 
\citep{sto78, chi86, cas90, lan91, bon00, mae00, mey02, kar03, pie04,pie06, val09,   
cha10, cas11, nei11, bro11a,bro11b} and empirical 
\citep{alc99, tes99, broc03, hun09a,hun09b, bona09,bona10, san09, efr11} investigations.    

One of the most interesting problem concerning evolutionary and pulsation 
properties of classical Cepheids is the so-called mass discrepancy problem. 
During the late sixties it was noticed by \citet{chr66,chr70} and by \citet{sto69} 
that the evolutionary masses --estimated using the comparison between isochrones 
and observations in the Color-Magnitude Diagram (CMD)-- were almost a factor of 
two larger than the pulsation masses --estimated using the 
Period-Mass-Radius relation-- of Galactic Cepheids \citep{fri71}. 

This conundrum was partially solved by \citet{mos92}  
using the sets of radiative opacities released by the OPAL \citep{igl91} 
and by the Opacity Project \citep{sea94} groups. 
However, several investigations focussed on Galactic \citep{bon01, cap05} 
and Magellanic \citep{bea01, bon02, kel02,kel06, kel08} Cepheids 
suggested that such a discrepancy was still of the 
order of 10-20\%. A similar discrepancy was also found by \citet{eva05a} 
using dynamical mass estimates of Galactic binary Cepheids and by 
\citet{broc03} using cluster Cepheids located in NGC~1866,  
a young Large Magellanic Cloud (LMC) cluster.
  
The Cepheid mass discrepancy problem can be addressed following two
different paths. \\ 

{\em a)} --{\em He-core mass}-- current evolutionary
predictions underestimate the He-core mass for intermediate-mass structures,
and in turn their Mass-Luminosity (ML) ratio. This possible drawback has
a substantial impact both on evolutionary and pulsation predictions, since
the latter models assume an ML relation. 

{\em b)} --{\em Envelope mass}--  Current Cepheid masses are smaller than
their main sequence (MS) progenitors, because they have lost a fraction of
their initial envelope mass. The latter working hypothesis implies that the
Cepheid mass discrepancy is intrinsic, i.e., it is not caused by limits in
the physical assumptions adopted in constructing evolutionary and pulsation
models. The input physics adopted in constructing evolutionary models of
intermediate-mass stars has already been addressed in several papers
\citep{chi86, broc93, sto93,sto96, cas04, ber09, val09}.

Here, we briefly mention the most relevant ones affecting the He-core mass 
and the envelope mass.

\centerline{\em a) He-core mass}   
{\em i)} --Extra-mixing-- Several detailed studies based on the comparison between 
predicted CMDs and Luminosity Functions of NGC1866, reached opposite conclusions in 
favor \citep{bar02} and against \citep{tes99, broc03} 
the occurrence of mild convective core overshooting. The need for a mild overshooting 
was also suggested by \citet{kel06} who investigated several young clusters 
in the LMC and in the Small Magellanic Cloud. More recently, it has also been 
suggested by \citet{cor02} that the degree of overshooting might also depend 
on the metal abundance, namely it increases when metal abundance decreases. 
{\em ii)} --Rotation-- 
Evolutionary models constructed by accounting for the effects of rotation
predict both an increase in Helium core mass and an enhancement in the surface
abundance of both Helium and Nitrogen \citep{mey02}. 
It has also been suggested that rotation might 
account for the significant changes in surface chemical compositions observed in 
Galactic and Magellanic supergiants \citep{kor05, hun09a,hun09b}. Moreover, recent 
theoretical \citep{mae00} and empirical \citep{ven99} investigations indicate that 
the efficiency of such a mechanism might depend on the initial metal abundance, i.e., 
it increases when metal abundance decreases. 
{\em iii)} --Radiative Opacity-- A new set of radiative opacities has been recently 
computed by the Opacity Project \citep{bad05}. The difference between old 
and new opacities is at most of the order of 5-10\% across the Z-bump (T$\approx$ 
250,000 K). To account for the mass discrepancy the increase in the opacity should 
be almost a factor of two. This indicates that the adopted radiative opacities have 
a marginal impact on the mass discrepancy problem.\\

\centerline{\em b) Envelope mass}  

{\em i)} --Canonical Mass Loss-- 
The total mass of an actual Cepheid is also affected by a decrease in the envelope 
mass. Evolutionary models accounting for the mass loss, during Hydrogen and Helium 
burning phases, by means of several semi-empirical relations \citep{rei75, nie90} 
do not solve the Cepheid mass discrepancy 
problem. Plausible values for the free parameter --$\eta$-- give mass loss rates 
that are too small. This is not surprising, since current semi-empirical 
relations are only based on scaling arguments and they are not rooted on a 
robust physical basis \citep{sch05}.  Empirical mass loss estimates based on 
NIR and ultraviolet emission for Galactic \citep{mca86, dea88, mar10a} and 
Large Magellanic Cloud \citep{nei08,nei09} Cepheids cover a broad range 
(10$^{-10}$ -- 10$^{-7}$ M$_\odot$ yr$^{-1}$). 
{\em ii)} --Enhanced Mass Loss-- 
More recent mid-infrared observations for 29 Galactic 
Cepheids collected with Spitzer indicate the presence of extended emission 
around seven of them \citep{barm11}. This finding together with 
the interferometric detection of circumstellar shells around several Galactic 
Cepheids \citep{mer06,mer07, ker06,ker08,ker09} further supports the evidence 
that there is a wind associated with Cepheids possibly triggered by 
pulsation-driven mass loss.     

A new spin concerning the occurrence of circumstellar shells around Cepheids 
was recently given by the detection of a large nebula around $\delta$ Cephei  
\citep{mar10b}, the protype of Cepheid variables. The IR emission 
was detected in several MIR bands and shows a radial extent larger than 
$10^4$ AU (5 arcmin). The nebula shows a parabolic shape and it is aligned 
with the direction of the motion, thus suggesting the possible occurrence 
of a bow shock caused by the interaction between the mass-losing Cepheid with 
the interstellar material. Current estimates indicate that $\delta$ Cephei 
may be losing mass  with a rate of $\sim$5$\times$10$^{-9}$ to 
$\sim$6$\times$10$^{-8}$ $M_\odot$ yr$^{-1}$. This finding has been independently
confirmed by \citet{mat11} using data collected in the radio with the Extended 
Very Large Array. They found an outflow velocity of $\sim$ 36 km/s and a mass 
loss rate of $\sim$ 10$^{-7}$--10$^{-6}$ M$_\odot$ yr$^{-1}$. 
The modest dust content of the outflow indicates that the wind from $\delta$ Cephei
might be pulsation-driven instead of dust-driven as typical for other classes of 
evolved stars. 

However, the stepping stone concerning the evolutionary and pulsation
properties of classical Cepheids was the detection of a Cepheid 
in a well detached,  
double-lined eclipsing binary system in the Large Magellanic Cloud. The geometry and 
the precision of both photometric and spectroscopic data gave the opportunity to 
measure the mass of the Cepheid (CEP0227) with the unprecedented precision of 1\% 
\citep[][hereinafter Paper I]{pietr10}. A similar precision was also attained 
for the classical Cepheid in the LMC eclipsing binary CEP1812 by 
\citep[][hereinafter Paper II]{pietr11}.  

This finding paved the road for new constraints on pulsation and evolutionary masses. 
Soon after, \citet{cas11} by using evolutionary models that account for a moderate 
convective core overshooting found evolutionary masses agree quite well with the dynamical 
mass of CEP0227 both in the Radius-Age plane and in the Radius-Effective Temperature plane.     
More recently, \citet{nei11} found that a combination of both moderate convective core 
overshooting and pulsation-driven mass loss is required to solve the Cepheid mass discrepancy. 

This is the third paper of a series focussed on the evolutionary and pulsation 
properties of Cepheids. The new series present two novel approaches when compared 
with similar investigations available in the literature:
{\em i)}-- the intrinsic parameters will be estimated using a Bayesian approach in dealing 
with the comparison between theory and observations;
{\em ii)}-- the empirical measurements are based on a homogeneous approach for both photometry 
and spectroscopy. In this investigation we provide new evolutionary constraints on 
the binary Cepheid CEP0227. 

The plan of the paper is the following. In \S2 we introduce the input physics adopted 
to construct evolutionary models. We also discuss the physical assumptions 
adopted to deal with mass loss, mixing and convective transport. In this section, we also 
outline the assumptions adopted to compute stellar isochrones.   
The description of the method adopted to estimate the intrinsic parameters of the binary 
components is given in \S 3. In this section we discuss the different most probable solutions 
in the L--T, in the M--R and in the g--T planes. \S 4 deals with the new empirical constraints 
on both stellar mass and age we obtained in the different planes for the binary system.  
In \S 5 we address the comparison between the most probable stellar isochrone and observations  
in the Color-Magnitude diagram. The true distance modulus and the reddening for the combined 
system are also discussed. In \S 6 we summarize the results and outline future possible 
avenues of the project.

\section{The theoretical framework}

\subsection{Input physics}
The evolutionary tracks were computed with an updated version of the FRANEC 
evolutionary code \citep{deg08, val09, tog11, dell11}. For these models, the 
2006 release of the OPAL equation of state (EOS) was adopted \citep{rog96}, 
together with the radiative opacity tables released in 2005 by the same Livermore 
group  \citep{igl96}\footnote{http://opalopacity.llnl.gov/opal.html.} for 
$\log T[K] > 4.5$ and extended at lower temperatures with the radiative opacity 
tables by \citet{fer05}. 
The conductive opacities are from \citet[][see also \citet{pot99}]{sht06}.
All the opacity tables were calculated for the \citet[][hereafter As09]{asp09}
heavy-element solar mixture.
The nuclear reaction rates are from the NACRE compilation 
\citep{nacre}, except for the $^{14}$N(p,$\gamma$)$^{15}$O, which is from the LUNA 
collaboration \citep{imb05,bem06,lem06} and 
$^{12}$C($\alpha$,$\gamma$)$^{16}$O from \citet{ham05}.  
These two reactions \citep[see, e.g.,][and references therein]{val09}  
affect both the extension and the morphology of the loop performed by 
intermediate-mass He-burning stars in the HR diagram. Hence the most 
recent values should always be adopted in stellar evolutionary codes 
in order to provide reliable theoretical predictions.   

Current calculations did not take into account He and metal diffusion, since its 
effect on the evolutionary properties of intermediate-mass stars is minimal. 
To model super-adiabatic convection, we adopted the mixing-length formalism 
\citep{boh58}, in which the efficiency of the convective transport depends 
on the mixing-length, i.e. $l=\alpha H_p$, where $H_p$ is the pressure height-scale 
and $\alpha$ is a free parameter. The $\alpha$ value --1.74-- was calibrated using the 
Standard Solar Models computed with both the same code and the same input physics, 
in particular using the same heavy-element mixture (As09), 
To account for current theoretical and empirical uncertainties, additional models 
with $\alpha$=1.9 were also computed. We also account for convective core overshooting, 
during central hydrogen burning phases, 
by extending the central mixed region of $l_{ov}$=$\beta H_p$ from the
border of the standard convective core defined by the Schwarzschild criterion, where 
$\beta$ is a free parameter \citep{cas00,val09}.
Note that we do not account for a mechanical core overshooting during the
central He-burning phase. The use of the Schwarzschild criterion to determine
the convective core extension, during these phases, would lead to an unphysical
discontinuity in the radiative gradient at the border of the convective core.
This effect is caused by the increase in the opacity due to the conversion of
He into C and O \citep{cas04}. To overcome the problem we model the growth
of the convective core and the development of a semiconvective region
following \citet{castellani71a,castellani71b}. Note that we neglect the
convective overshooting at the base of the convective envelope during shell
hydrogen burning phases.  
   
\begin{figure*}[th]
\includegraphics[width=0.7\textwidth,angle=270]{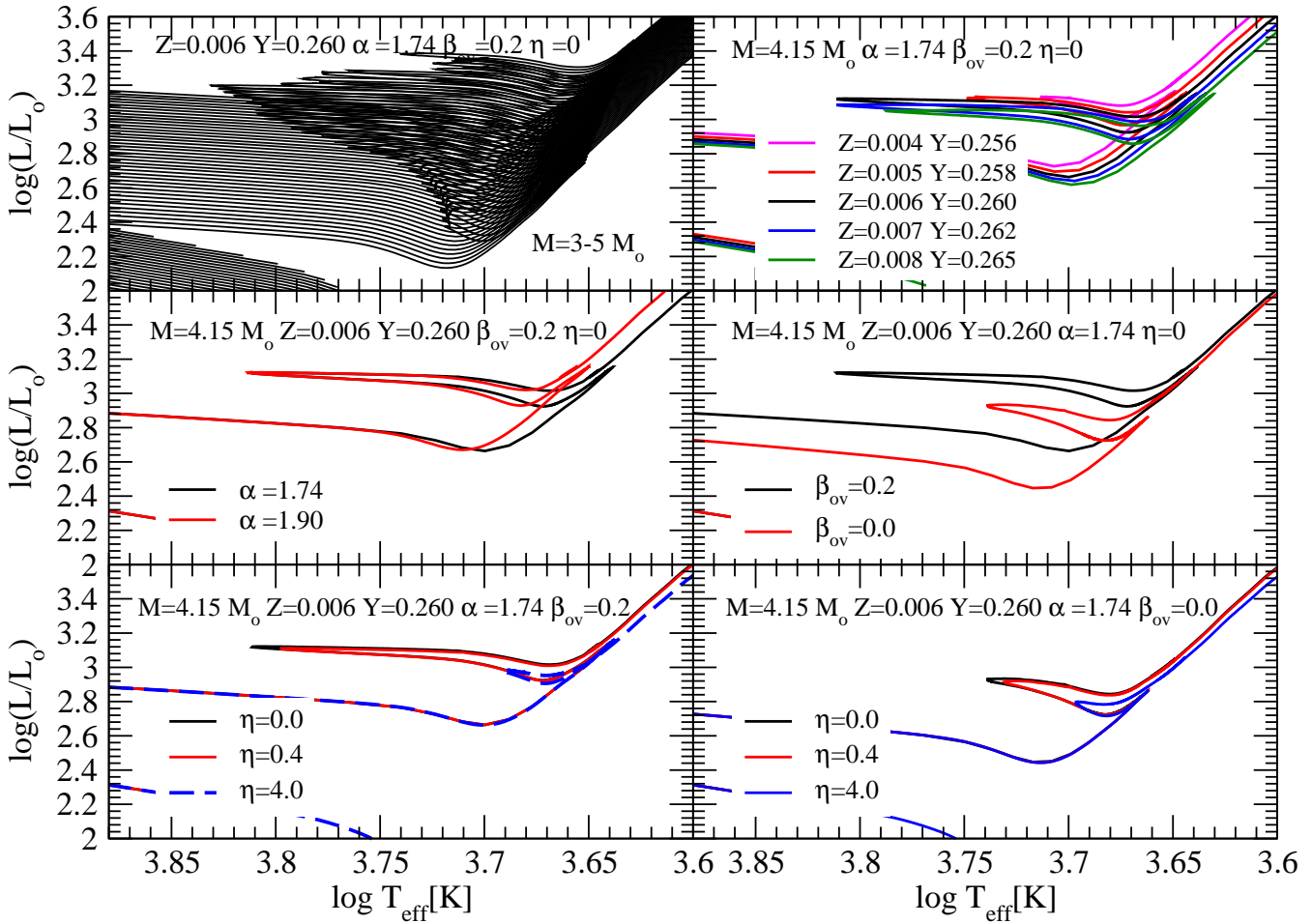}
\label{plotP}
\caption{Top -- Left -- Hertzsprung-Russell Diagram of a set of evolutionary tracks  
computed at fixed metal (Z=0.006) and helium (Y=0.260) abundance and stellar masses 
ranging from 3 to 5 M$_\odot$. 
The plot is focussed around the evolutionary phases of central helium burning 
(blue loops). The free parameters adopted to deal with 
convective transport (mixing length, $\alpha$), with convective core 
overshooting ($\beta$) and with mass loss ($\eta$) are also labeled.    
Top -- Right -- Same as the Top--Left, but for evolutionary tracks computed at fixed 
stellar mass (M=4.15 M$_\odot$), but different chemical compositions (see labeled 
values).  
Middle -- Left -- Same as the Top--Left, but for evolutionary tracks computed at fixed 
stellar mass (M=4.15 M$_\odot$) and chemical composition (Z=0.006, Y=0.260), but for 
two different assumptions concerning the mixing length (see labeled values). 
Middle -- Right -- Same as the Top--Left, but for evolutionary tracks computed at fixed
stellar mass (M=4.15 M$_\odot$) and chemical composition, but for two different 
assumptions concerning the convective core overshooting.  
Bottom -- Left -- Same as the Top--Left, but for evolutionary tracks computed at fixed
stellar mass (M=4.15 M$_\odot$) and chemical composition, but for three different 
assumptions concerning the mass loss rate. 
Bottom -- Right -- Same as the Bottom--Left, but for evolutionary tracks computed 
neglecting the convective core overshooting.  
}
\end{figure*}

\begin{figure*}
\includegraphics[width=0.99\textwidth]{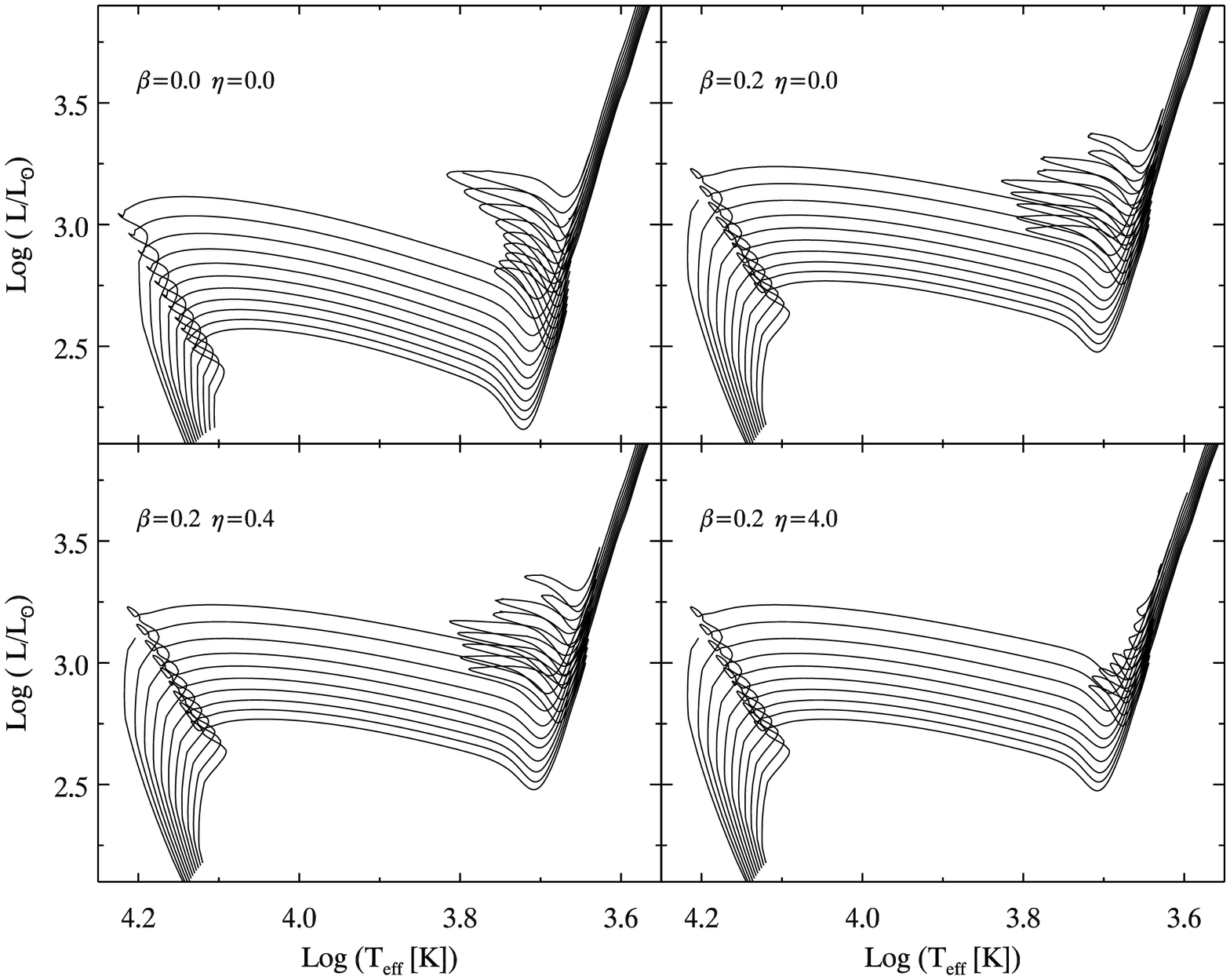}
\label{fig:ploISO}
\caption{Hertzsprung-Russell Diagram of sets of isochrones  
computed at fixed metal (Z=0.006) and helium (Y=0.260) abundance and ages ranging 
from 80 to 180 Myr, with a step in age of 10 Myr.}
\end{figure*}

\subsection{Evolutionary tracks}

The approach we plan to adopt to estimate the intrinsic parameters of the binary components 
does require detailed set of evolutionary tracks and of isochrones properly covering the 
helium burning phases. 
The stellar masses of the evolutionary tracks range from 3 to 5 $M_\odot$, and the step 
in stellar mass is 0.05 $M_\odot$. Therefore, each set includes 42 evolutionary tracks.  
Unfortunately, we still lack an accurate measurement of the iron and heavy element 
abundances of CEP0227. Therefore, we adopted the mean metallicity of LMC Cepheids 
([Fe/H]=-0.33$\pm$0.13 dex) based on recent high-resolution, high signal-to-noise 
spectra collected with UVES at ESO/VLT \citep{roma08}.          
The top, left panel of Fig.~1 shows the set of evolutionary models computed at fixed 
helium (Y=0.260) and metal (Z=0.006) content and stellar masses ranging from 3 to 5 $M_\odot$. 
The evolutionary phases plotted in this panel cover the central helium burning evolutionary 
phases (blue loops). 

To take into account possible changes in iron abundance, due to intrinsic metallicity 
dispersion \citep{pom08, roma08, hun09a, hun09b} we also computed four independent 
sets of evolutionary tracks with metallicity ranging from Z=0.004 to Z=0.008, 
while the helium content was fixed assuming an helium-to-metal enrichment ratio
$\Delta Y$/$\Delta Z$=2 \citep[][]{pag98, jim03, fly04, gen10} 
and a primordial helium abundance of $Y_p$=0.2485 
\citep[see, e.g.,][]{cyb05, ste06, pei07, kom11}. Therefore, current helium values 
range from Y=0.256 to Y=0.265 (see top, right panel of Fig.~1).   

We already mentioned that intermediate-mass stars are affected by uncertainties on  
the actual size of the He-core mass. To constrain the impact of the convective core 
overshooting during central hydrogen burning phases on the intrinsic parameters, the 
evolutionary tracks were computed either neglecting ($\beta_{ov}$=0) or accounting 
for a moderate convective core overshooting ($\beta_{ov}$=0.2, see top panels of Fig.~1).  
The two components of the binary system are relatively cool objects (see Table~1) with 
extended convective envelopes. To constrain the dependence on the efficiency of convective 
transport in the super-adiabatic regions the calculations were performed assuming two 
different mixing length parameters. 
The former one --$\alpha$=1.74-- calibrated using the standard solar model 
\citep[][]{dell11} and the latter one --$\alpha$=1.90-- to mimic a more efficient 
convection (see middle panels of Fig.~1).

Recent theoretical investigations based on evolutionary models accounting for both 
convective core overshooting and for a pulsation driven mass loss suggested that the 
latter physical mechanism can play a crucial role in settling the Cepheid mass discrepancy 
problem \citep{nei11}. A detailed analysis of the dependence of the blue loops 
on this physical mechanism is out of the aim of this investigation. However, to constrain 
the dependence of the most probable solution on the mass loss rate, different sets of 
evolutionary models were computed accounting for mass loss after the central hydrogen
exhaustion. We used the Reimers formula

$dM/dt$ = $\eta$ $\times$4 $\times$ $10^{-13}$ $\times$ L/gR ~~~~~~[$M_\odot$/yr]

where $\eta$ is a free parameter, L,R and g are the luminosity, the radius, 
and the surface gravity in solar units. Sets of models with four different 
$\eta$ values, namely 0, 0.4, 0.8 and 4.0, were computed. The evolutionary 
tracks constructed by assuming $\eta$=0.4 and 0.8 show marginal changes 
and in the following we will only take into account the former ones 
(see bottom panels of Fig.~1). 
 
We ended up with a sample of more than 2,000 evolutionary tracks computed 
for this specific project.

\subsection{Stellar isochrones}

For each set of evolutionary tracks discussed in \S 2.2, we also computed a fine grid 
of stellar isochrones with ages ranging from a few tens to a few hundreds of Myr. 
To improve the accuracy of the parameter estimates (see \S \ref{sec:parest}) and to 
provide robust determinations of the confidence interval, the isochrones were further 
interpolated. The actual grid of isochrones has a step in age of 0.02 Myr, necessary to 
cover the fast evolution of the stars across the blue loop.

Subsets of isochrones for selected input parameters (see labeled values) in the 
HR diagram are shown in Fig.\ref{fig:ploISO}.

\section{Parameter estimates}
\label{sec:parest}
The parameters that determine the evolutionary status of a star are its mass, 
$\mu$, its age, $\tau$ and its chemical composition, $\zeta$.
The value of these parameters determine the position of a star in the HR diagram or, 
similarly, in any diagram were observable properties are displayed.
Stellar models predict observables as a function of these three parameters, 
hence their values can be estimated by comparing predictions and observations.
To provide robust estimates of both stellar masses and ages of the binary components 
we followed the Bayesian approach described in \citet{gen12}. The interested 
reader is referred to the quoted paper, here we only summarize the relevant points of 
the approach we adopted.

\subsection{Bayesian approach}

Given a set of observables $\vec{q}$ the probability of the model parameters, 
$\vec{p} = (\mu,\tau,\zeta)$ is given by:

\begin{equation}
f( \vec{p} | \vec{q} ) \propto \mathcal{L}(\vec{p}|\vec{q}) f(\vec{p}) .
\end{equation}

Here $\mathcal{L}(\vec{p}|\vec{q})$ is the Likelihood of the parameters $\vec{p}$ 
given the set of observations $\vec{q}$ and $f(\vec{p})$ is the prior 
distribution of the parameters.

We used three different pairs of observables, $\vec{q}$, to determine the 
$(\mu,\tau,\zeta)$ set of parameters of the binary components. In particular, 
we performed the comparison in the $\vec{q} = (\log L/L_\odot, \log T_{\mathrm{eff}}), 
(M/M_\odot, R/R_\odot)$ and $(\log T_{\mathrm{eff}}, \log g)$ planes.

To estimate the best values of the stellar masses and ages the probability 
$f( \vec{p} | \vec{q} )$ is marginalized w.r.t the other two variables:
\begin{eqnarray}
\label{eq:Gtau}
 G(\tau) & = & \int \mathcal{L}(\tau,\mu,\zeta|\vec{q})  f(\tau,\mu,\zeta)\, \mathrm{d}\mu \,\mathrm{d}\zeta \\
\label{eq:Hmu}
H(\mu)  & = & \int \mathcal{L}(\tau,\mu,\zeta|\vec{q})  f(\tau,\mu,\zeta)\, \mathrm{d}\tau \, \mathrm{d}\zeta
\end{eqnarray}

The mode of the marginal distributions is used as the best parameter estimator 
\citep{jor05, gen12}. 
Confidence intervals are defined by excluding 16\% of the area under the posterior 
distribution on each side of the variable domains.

In our treatment the metallicity is not considered as an unknown parameter to be determined. 
On the contrary, we adopted the available prior information on the metallicity of Cepheids 
in the LMC to constrain our predictions on the stellar masses and ages.
In detail, we adopted a Gaussian distribution of [Fe/H] with mean $<\mathrm{[Fe/H]}> = -0.33$ 
and standard deviation $\sigma_{\mathrm{[Fe/H]}} = 0.13$ dex. 
The quoted values are based on the metallicity distribution of LMC Cepheids measured by 
\citet{roma08}. 

Given the available information on the dynamical stellar masses we adopted two different priors 
for the stellar mass. In one case a flat prior is adopted, defined across the entire range of 
stellar masses for which we computed evolutionary tracks: $M\in[3,5]M_{\odot}$. In the second 
case we used a prior with a Gaussian distribution with mean and standard deviations based on 
the dynamical mass estimates by \citet{pietr10}.

Stellar ages were estimated separately for the two binary components and for the system as a 
whole. The age of each star is determined from the $G(\tau)$ marginal distribution of equation 
(\ref{eq:Gtau}). 
In the reasonable hypothesis that the two components of the binary system have the same age, 
their composite age distribution, i.e. the age of the system, can be simply derived  as:
\begin{equation}
 G_{\mathrm{C}} = G_{\mathrm{P}} \times G_{\mathrm{S}} \; ;
\end{equation}
where the subscripts indicate the composite (C), the primary component (P) and the 
secondary component (S) age distributions. In a recent investigation, 
\citet{gen12} showed that the age distribution of a stellar system 
based on this approach is a more precise indicator of the real age when 
compared to single stellar ages.

The above approach was applied to each set of evolutionary models listed in Table~1, i.e. 
for each different assumption concerning the value of the overshooting parameter 
--$\beta$-- the mass-loss --$\eta$-- and the mixing length --$\alpha$-- parameter. 
The posterior probabilities were computed using both the flat and the Gaussian prior on the 
stellar mass. Selected results of the individual solutions are given in Table~3. 
The solutions based on different observables are discussed in the following   
subsections.

\subsection{The Hertzsprung-Russell Diagram}

As a first step to constrain the intrinsic parameters of the binary system, we 
applied our method in the Hertzsprung-Russell Diagram. The main 
advantage in using this plane is that we can estimate the intrinsic luminosities 
using the Stefan-Boltzmann relation. 
Therefore, the solution is independent of distance modulus, reddening and stellar 
mass. The main drawback is that the two parameters (luminosity and temperature) 
are correlated and more affected by measurement errors than the stellar mass 
and the radius (see Fig.~3).  
However, our method takes into account the fact that temperature and luminosity are
not independent, and indeed the covariance matrix is also included in the solution  
\citep[see equations (3), (4) and (5) in][]{gen12}.

\begin{figure*}
\includegraphics[width=0.99\textwidth]{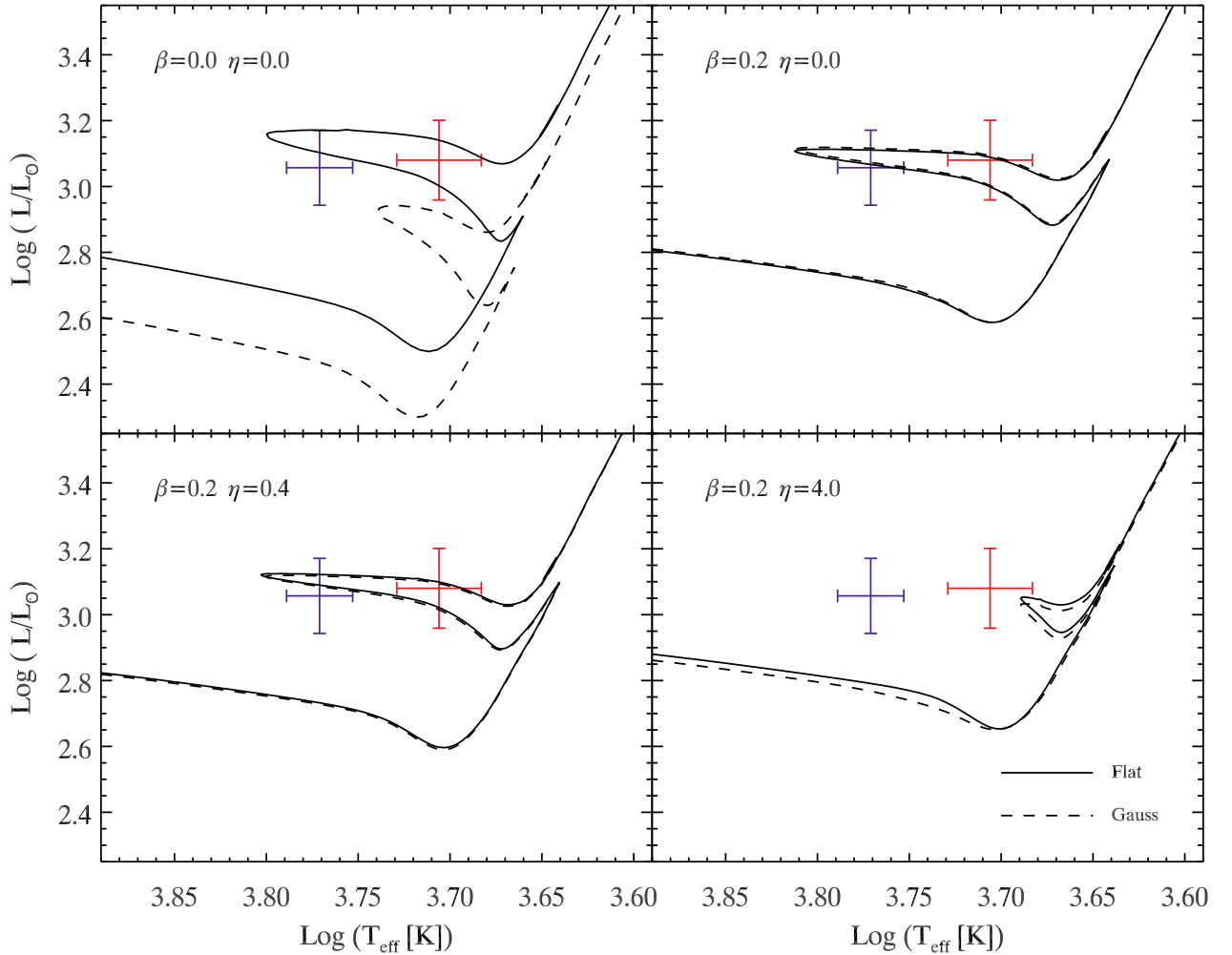}
\label{plotP}
\caption{Top -- Left-- Comparison in the Hertzsprung-Russell Diagram between the binary 
components and the isochrone giving the maximum of 
$G_{\mathrm{C}} = G_{\mathrm{P}} \times G_{\mathrm{S}}$ (see Table~3).
In this panel the solution is based on evolutionary models constructed by neglecting convective 
core overshooting ($\beta$=0) and mass loss ($\eta$=0). The solid and the dashed lines 
show the isochrones with most probable age derived using the Gaussian and the flat prior 
on the mass distribution, respectively. The blue bar marks the position of the Cepheid 
(primary), while the red ones the position of the red giant (secondary).       
The intrinsic parameters of the isochrones are also labeled.  
Top --Right-- Same as top left, but for 
models accounting for mild ($\beta$=0.2) convective core overshooting. 
Bottom -- Same as the top right, but for models accounting for a canonical 
(left, $\eta$=0.4) and an enhanced (right, $\eta$=4.0) mass loss rate.}. 
\end{figure*}

\subsection{The stellar mass vs radius plane}

To further constrain the intrinsic parameters of the binary system  we also applied 
the Bayesian method in the stellar mass vs radius plane. The main advantage in using 
this plane is that the adopted observables are independent of distance modulus and 
reddening and they are also independent from each other (see Fig.~4). 

\begin{figure*}
\includegraphics[width=0.99\textwidth]{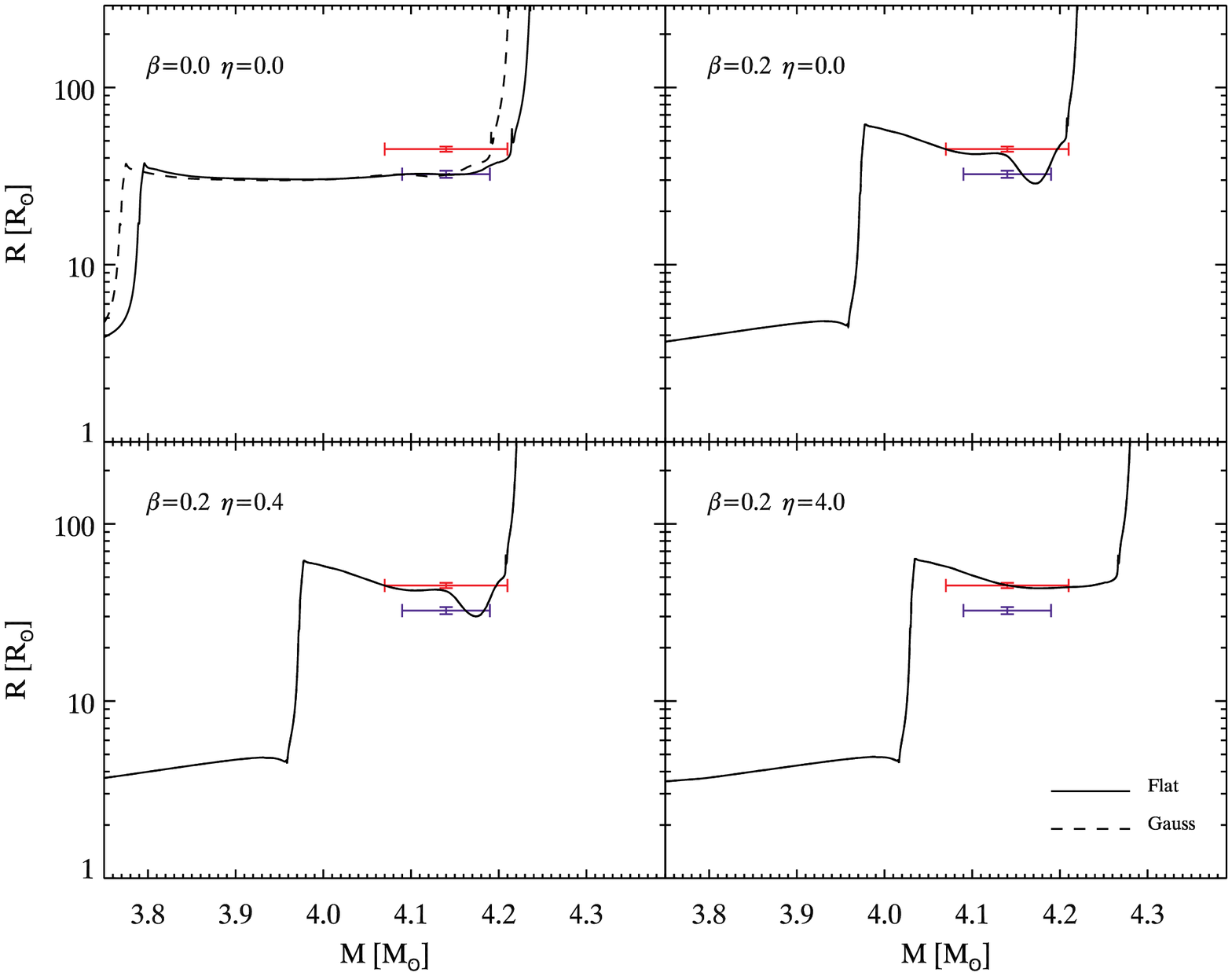}
\label{plotP}
\caption{
Top --Left-- Comparison in the stellar mass vs radius plane between the binary
components and the isochrone giving the maximum of
$G_{\mathrm{C}} = G_{\mathrm{P}} \times G_{\mathrm{S}}$ (see Table~3).
The symbols and the lines are the same as in Fig.~3.
Top --Right-- Same as top left, but for models accounting for mild ($\beta$=0.2)
convective core overshooting. 
Note that the dashed line overlaps with the solid line, since they attain very 
similar values in the M--R plane.
Bottom -- Same as the top right, but for models accounting for a canonical
($\eta$=0.4, left) and an enhanced ($\eta$=4.0, right) mass loss rate.}  
\end{figure*}

\subsection{The $\log T_{eff}$ vs $\log g$ plane}

\begin{figure*}
\includegraphics[width=0.99\textwidth]{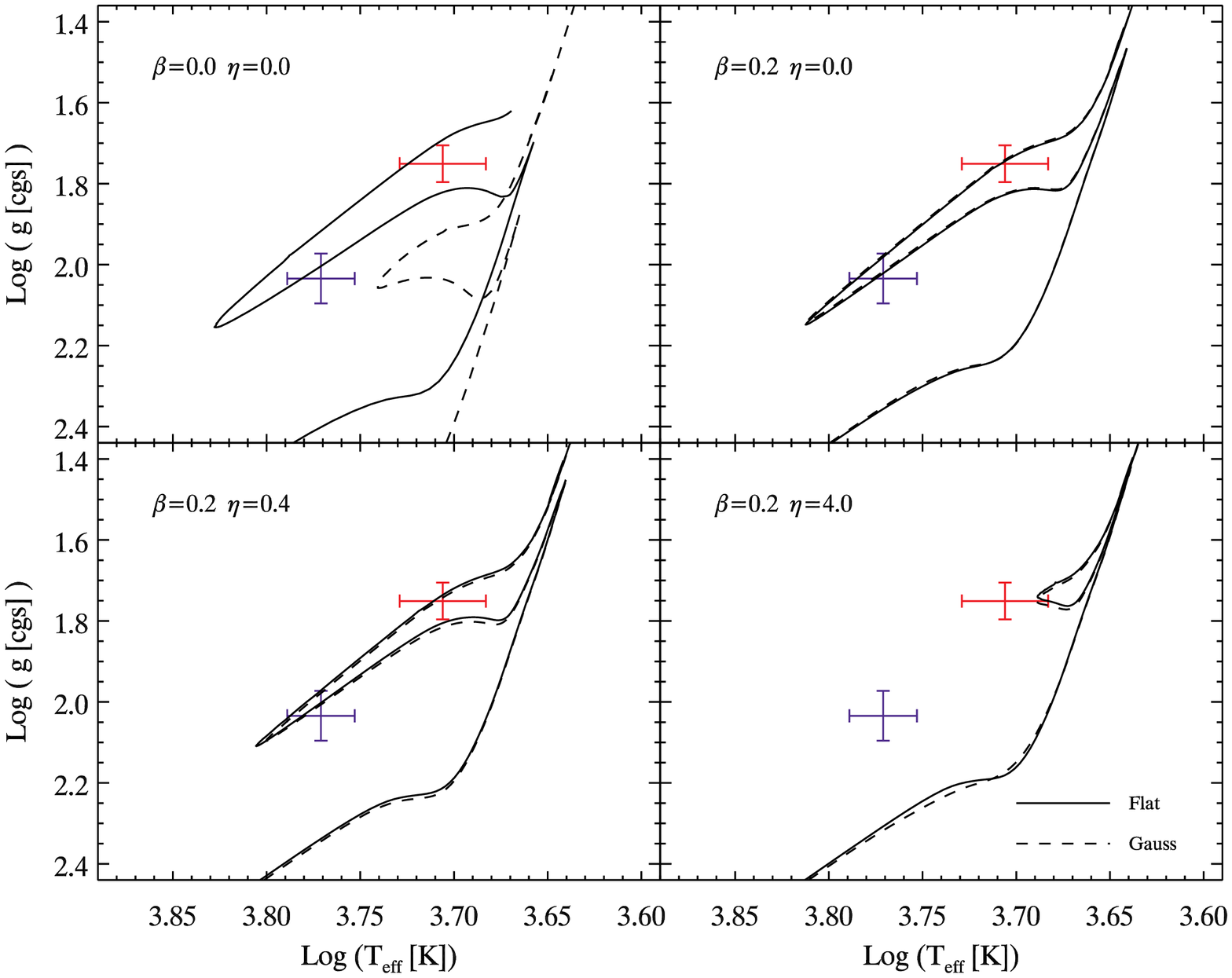}
\label{plotP}
\caption{Top --Left-- Comparison in the logarithmic effective temperature vs logarithmic 
surface gravity plane between the binary components and the isochrone giving the maximum 
of $G_{\mathrm{C}} = G_{\mathrm{P}} \times G_{\mathrm{S}}$ (see Table~3).
The symbols and the lines are the same as in Fig.~3. 
Top --Right-- Same as top left, but for models accounting for mild ($\beta$=0.2) 
convective core overshooting.  Bottom -- Same as the top right, but for 
models accounting for a canonical (left, $\eta$=0.4) and an enhanced (right, $\eta$=4.0) 
mass loss rate.}. 
\end{figure*}

Finally, we applied the Bayesian method also into the effective temperature vs surface 
gravity plane. The main advantage in using this plane is that the adopted observables 
are independent of distance modulus and reddening. Moreover, the observables are 
independent from each other and the isochrones display a relevant difference 
in this age range (see Fig.~5). This plane combines three independent measurements, 
thus providing the most stringent test for the theory.

\section{Constraints on the stellar mass and age of the binary system}

The most probable mass and age values that we found indicate that the mixing length 
parameter has a minimal impact on the estimate of these parameters. This result is 
not surprising, since the extent in effective temperature of the blue loop is minimally 
affected by the change in $\alpha$ from 1.74 to 1.90. 
This is the reason why in Table~3 we only included the results based on models with 
$\alpha$=1.74.  What is clear from the values listed in Table~3 is that the solutions 
based on evolutionary models that do not account for the convective core overshooting 
and those with enhanced mass-loss ($\eta=4.0$) are in worse agreement with the measured 
dynamical masses, when compared with the results based on models accounting for the 
convective core overshooting and $\eta=0.0$ or 0.4. 

In the case of a flat mass prior, the $\beta =0.0$ models predict stellar masses that 
are systematically larger than observed, even though the difference with the dynamical 
mass is typically within the 1$\sigma$ confidence interval. 
Similarly, the models with enhanced mass-loss, in the case of flat mass prior, tend
to underestimate the mass of the primary component. These two sets of models give a 
good agreement with the data, only when the method is used in the M--R plane.

Comparing the results for a flat and a Gaussian mass prior, it is clear that the 
latter drives the solution towards the measured dynamical mass. While this might 
appear obvious, we will show in Sect. \ref{sec:bf} that not only the mode of the 
mass distributions are shifted towards the measured values, but also the whole 
solution has a higher reliability, measured using the so-called $\mathrm{evidence}$ 
parameter (see \S 4.1). 
The impact of the Gaussian prior is very important also for the age results.
In the Gaussian mass prior case, the age solutions for the two components 
are generally more precise, meaning that the uncertainty interval is narrower 
compared to the flat prior case. In addition to this, there is also a better 
agreement between the ages of the two components, and therefore, a more 
robust result for the composite age of the system.

For each set of models, we plotted the isochrones corresponding to the most 
probable composite age (indicated by a C in Table~3) in Figs.~3, 4 and 5. 
The three figures show the most probable isochrone as obtained by applying 
the method in the L--T, in the M--R and in the g--T plane. Solid and dashed 
lines display the most probable ages for a flat and  a Gaussian mass 
prior, respectively.

In passing, we note that evolutionary models constructed assuming a slightly 
enhanced mass loss rate ($\eta$=0.8) also provide very similar solutions to 
the canonical ($\eta$=0.4) case. 
This indicates that canonical recipes to account for mass loss rate do not 
allow us to provide more quantitative constraints on mass loss efficiency 
during advanced evolutionary phases \citep{nei11}. However, a detailed treatment 
of the pulsation-driven mass loss inside the instability strip is out of the 
aim of this investigation.  

\subsection{$\mathrm{Evidence}$ of the most probable solutions}
\label{sec:bf}

The confidence intervals of the different solutions provide a hint concerning the 
robustness of individual fits using the different observables. 
However the precision of the result alone migth be a deceptive indicator of the 
goodness of the fit between theory and data.
A narrower confidence interval in the $(\mu,\tau)$ parameter space does not 
necessarily imply that the corresponding choice either of the observables, 
or of the mass prior or of the set of models is better than other possible choices.

A quantitative measurement of the relative goodness of the different solutions 
can be provided by the Bayes Factor.   
The Bayes Factor, $\mathrm{BF}_{ij}$, between two sets of models is defined 
as the ratio of the $\mathrm{evidence}$ of the two models. The $\mathrm{evidence}$ 
itself is the integral of the Likelihood, marginalized over the model parameters prior distribution 
\citep[see, e.g.,][]{bj11}. The $\mathrm{evidence}$ gives the total probability of the observed 
data given the particular set of models considered. Since our sets of models differ 
for the choice of the mixing length parameter, $\alpha$, the core overshooting 
parameter, $\beta$, and the mass-loss parameter, $\eta$, we introduce 
$\Xi_i = (\alpha_i,\beta_i,\eta_i)$ to identify a particular choice of 
the three parameters. The $\mathrm{evidence}$ is therefore:
\begin{equation}
\label{eq:evid}
\mathcal{E}_i(\vec{q}) \equiv \mathcal{E}(\vec{q}|\Xi_i) = \int \mathcal{L}(\tau,\mu,\zeta|\vec{q},\Xi) f(\tau,\mu,\zeta|\Xi) \mathrm{d}\tau \, \mathrm{d}\mu \, \mathrm{d}\zeta
\end{equation}

The $\mathrm{evidence}$ depends on the plane in which the comparison between data and 
models is performed (L--T, M--R, g--T), on the choice of the prior distribution 
of the model parameters (i.e., $f(\tau,\mu,\zeta|\Xi)$, in our case flat or 
Gaussian mass distribution and Gaussian metallicity distribution) and on the 
$\Xi$-set of parameters identifying that set of models (see Fig.~6).
$\mathcal{E}_i(\vec{q})$ is proportional to the total probability of observing 
the data $\vec{q}$ under the assumption that the $\Xi_i$ model is the correct model.

In equation (\ref{eq:evid}) the $\vec{q}$ observables refer to a single star. 
The composite $\mathrm{evidence}$ for both stars is simply:
\begin{equation}
 \mathcal{E}_i( \{ \vec{ q_{ \mathrm{P} } }, \vec{ q_{ \mathrm{S} } } \} ) = \mathcal{E}_i( \vec{ q_{\mathrm{P}} } ) \times \mathcal{E}_i( \vec{ q_{\mathrm{S}} } )
\end{equation}

From equation (\ref{eq:evid}), it follows that the Bayes factor is:
\begin{equation}
 \mathrm{BF}_{ij} = \frac{\mathcal{E}_i}{\mathcal{E}_j}
\end{equation}

The Bayes factor does not constrain the triplet $(\tau,\mu,\zeta)$ that provides the 
highest posterior probability for each set of models. However, it can be used to 
quantify which set of models, variable plane and prior distribution choice give 
the best overall agreement with the data.
Typically, in order to claim that a set is preferred over another, the Bayes factor 
should be smaller/larger than one, with either BF $< 0.1$ or BF $> 10$ \citep{kr95}.

Table~4 gives the BF values using as a reference the set of models with 
$\alpha=1.74$, $\beta=0.2$ and $\eta=0.4$, together with the variables 
$\vec{q} = (\log g, \log T_{\mathrm{eff}})$ and with the Gaussian mass 
prior distribution. This reference set gives the largest $\mathrm{evidence}$ for 
the composite system, therefore it can be considered --within our range of
models-- the best set of models and the best prior distribution choice to 
describe the observations.
The BFs listed in Table~4 were estimated by dividing the corresponding 
$\mathrm{evidence}$ by the $\mathrm{evidence}$ of the reference set either 
for the primary star (P entries), or for the secondary star (S entries) 
or for the composite $\mathrm{evidence}$ (C entries).

The ($\beta=0.2, \eta=0.0$) set of models gives BFs for the primary 
(see Fig.~6), the secondary (see Fig.~7) and the composite system that 
are very similar to the 
reference set. Therefore, these two sets give similar likely solutions 
and with the present data we can not distinguish between them. 
On the contrary, the set without overshooting, $\beta=0$, and the set 
with enhanced mass-loss rate, $\eta=4.0$, show very low values of the BFs, 
suggesting that these parameters can be excluded, hence, confirming the 
qualitative analysis at the beginning of this section.

It is worth noticing that, the BFs for a given $\vec{q}$ plane and 
a given $\Xi$ set are generally larger  when the Gaussian mass prior 
is used. This suggests that the likelihoods themselves are already 
centered in a region of the parameter space which is close to the 
dynamical mass measurements. Therefore, the Gaussian prior enhances 
the posterior probabilities when compared with the flat case. 
This finding is also supported by the fact that several solutions  
obtained using a flat mass prior distribution are characterized by large 
confidence intervals (see top panels in Figs.~6 and 7 and the values 
listed in Table~3). In addition to this, the plane in which the largest 
BFs are found is the g--T plane. This means that the models, when the 
measurements of stellar mass, radius and effective temperature are combined, 
are still able to reproduce these three independent observables, and actually 
the total probability increases. 

\begin{figure*}[!h]
\begin{center}$
\begin{array}{ccc}
\includegraphics[width=0.32\textwidth]{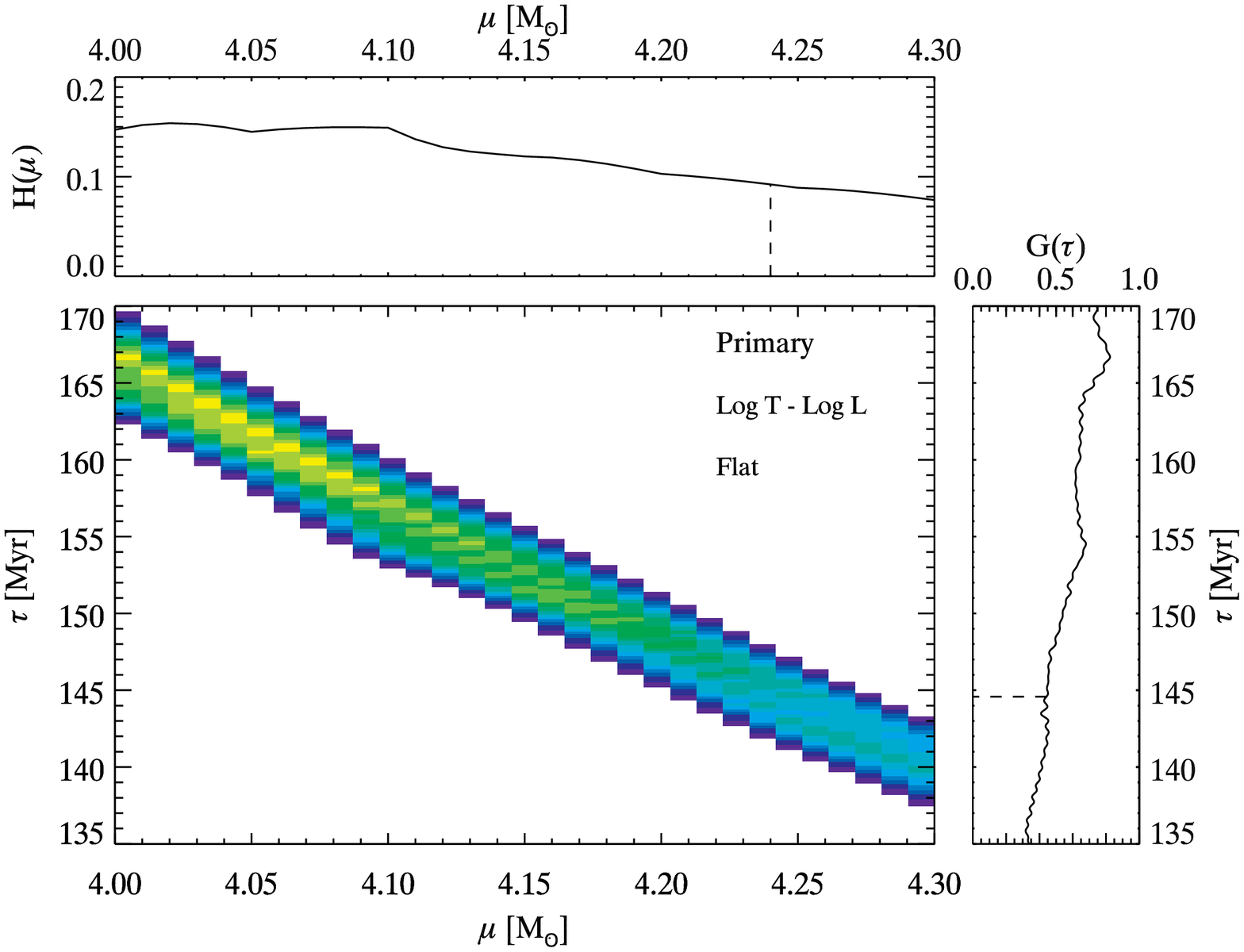} & 
\includegraphics[width=0.32\textwidth]{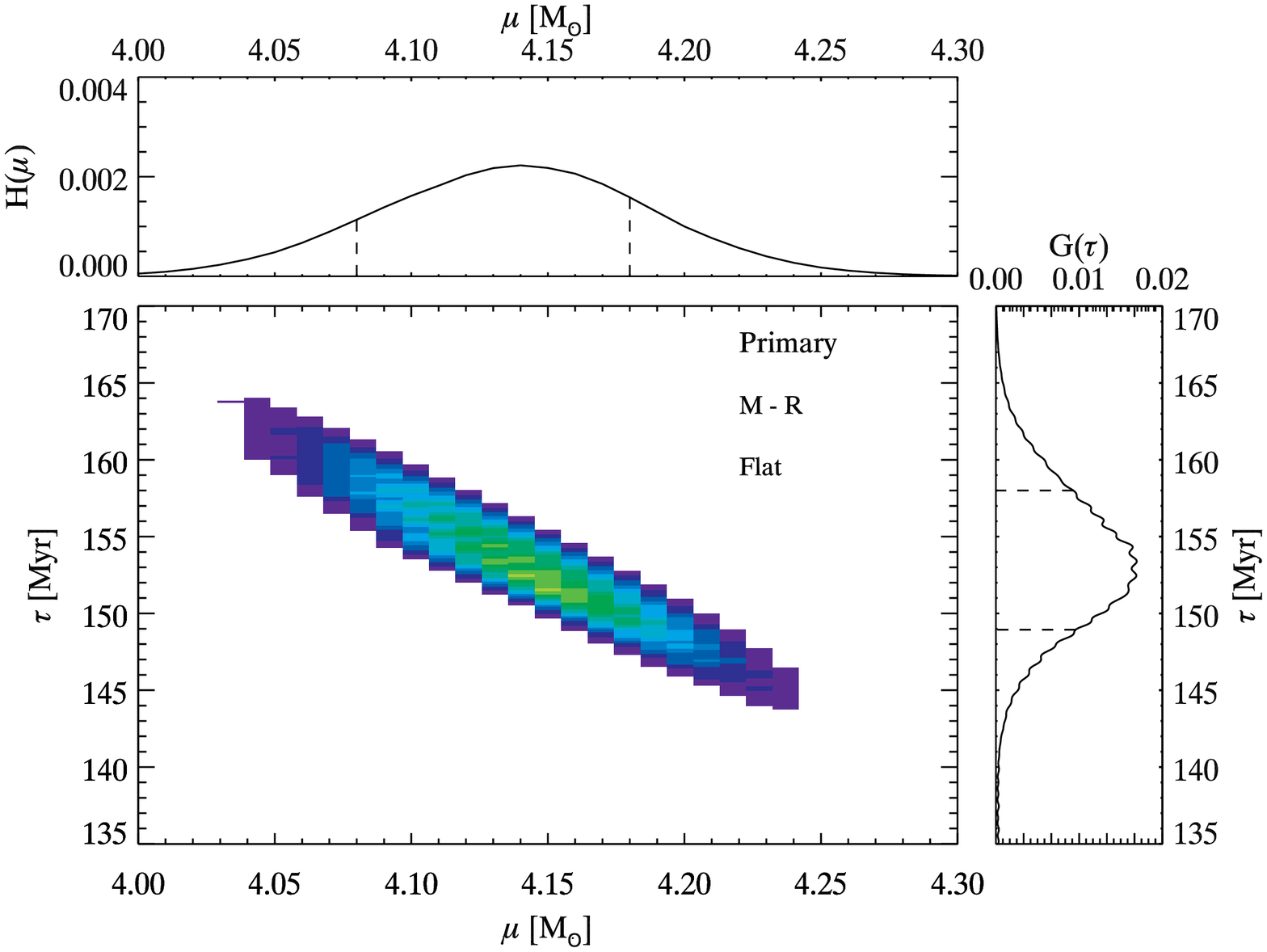} & 
\includegraphics[width=0.32\textwidth]{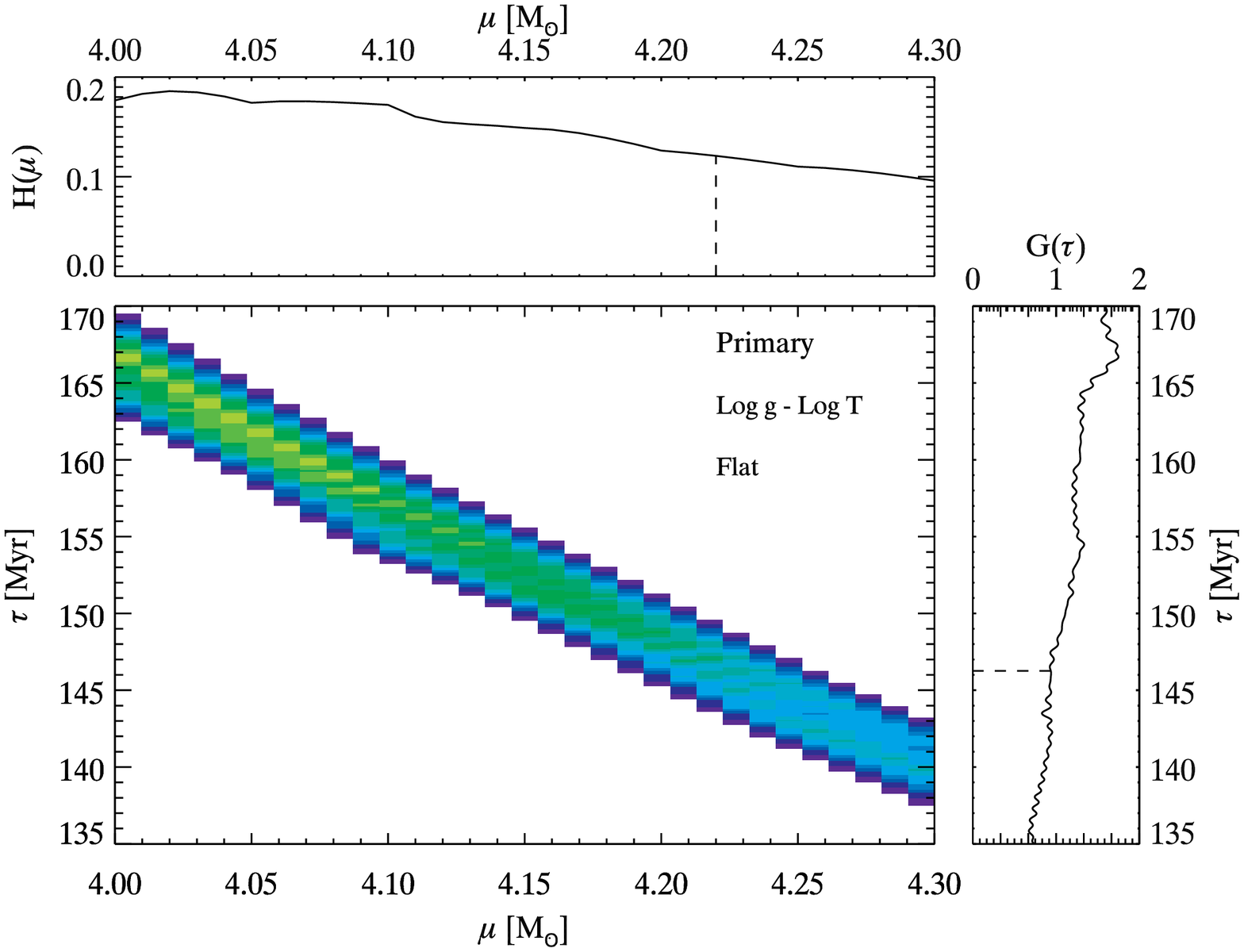} \\  
\includegraphics[width=0.32\textwidth]{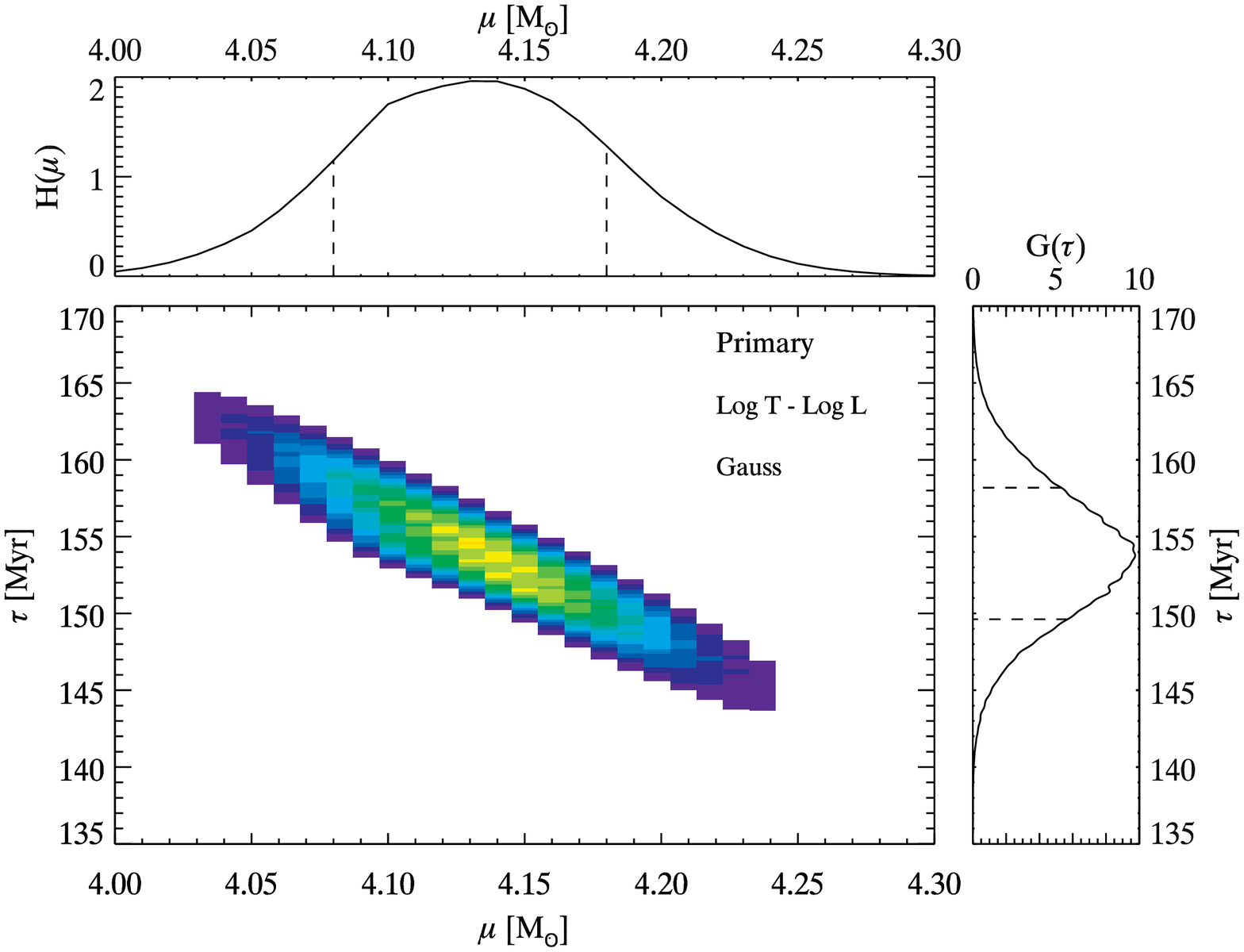} & 
\includegraphics[width=0.32\textwidth]{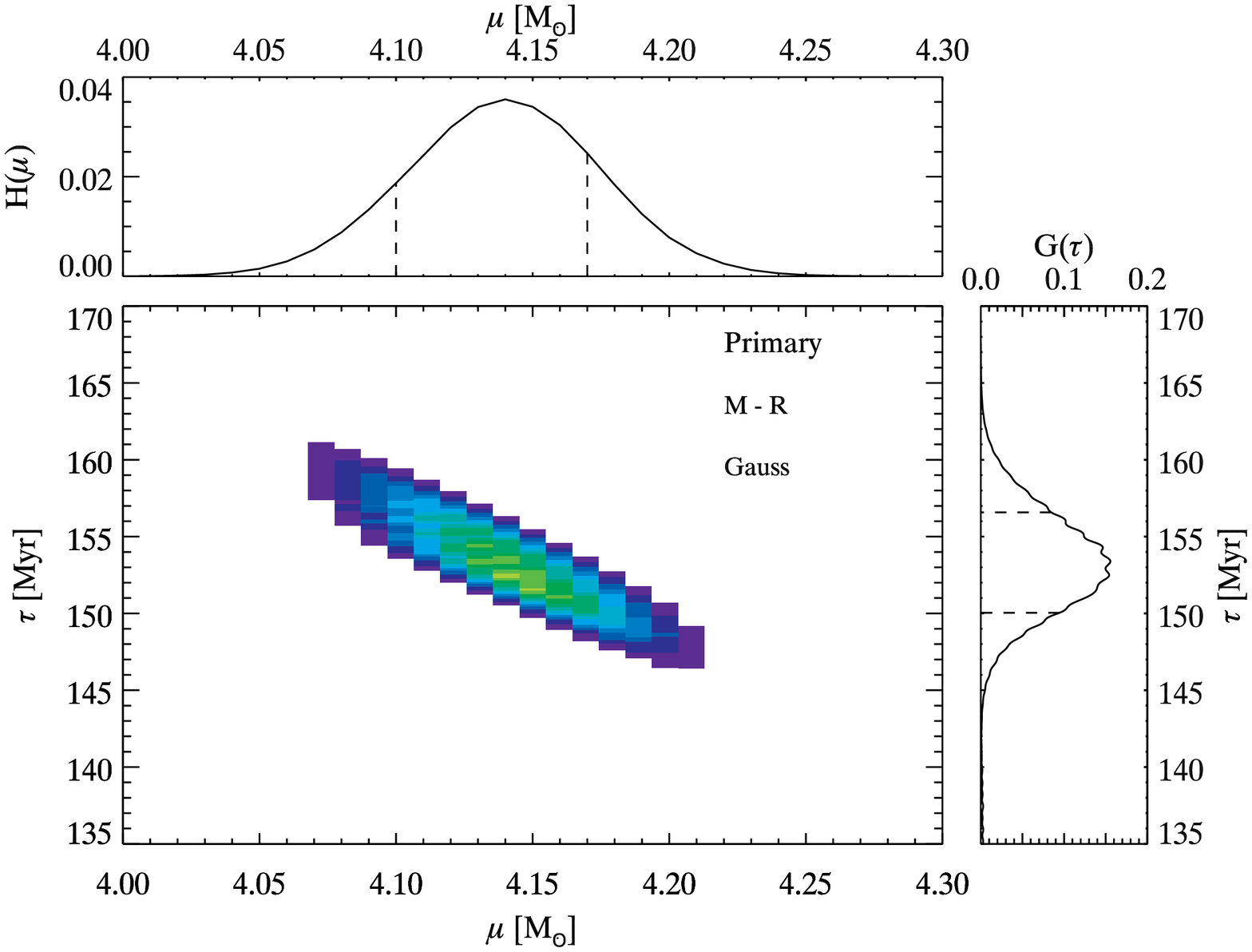} & 
\includegraphics[width=0.32\textwidth]{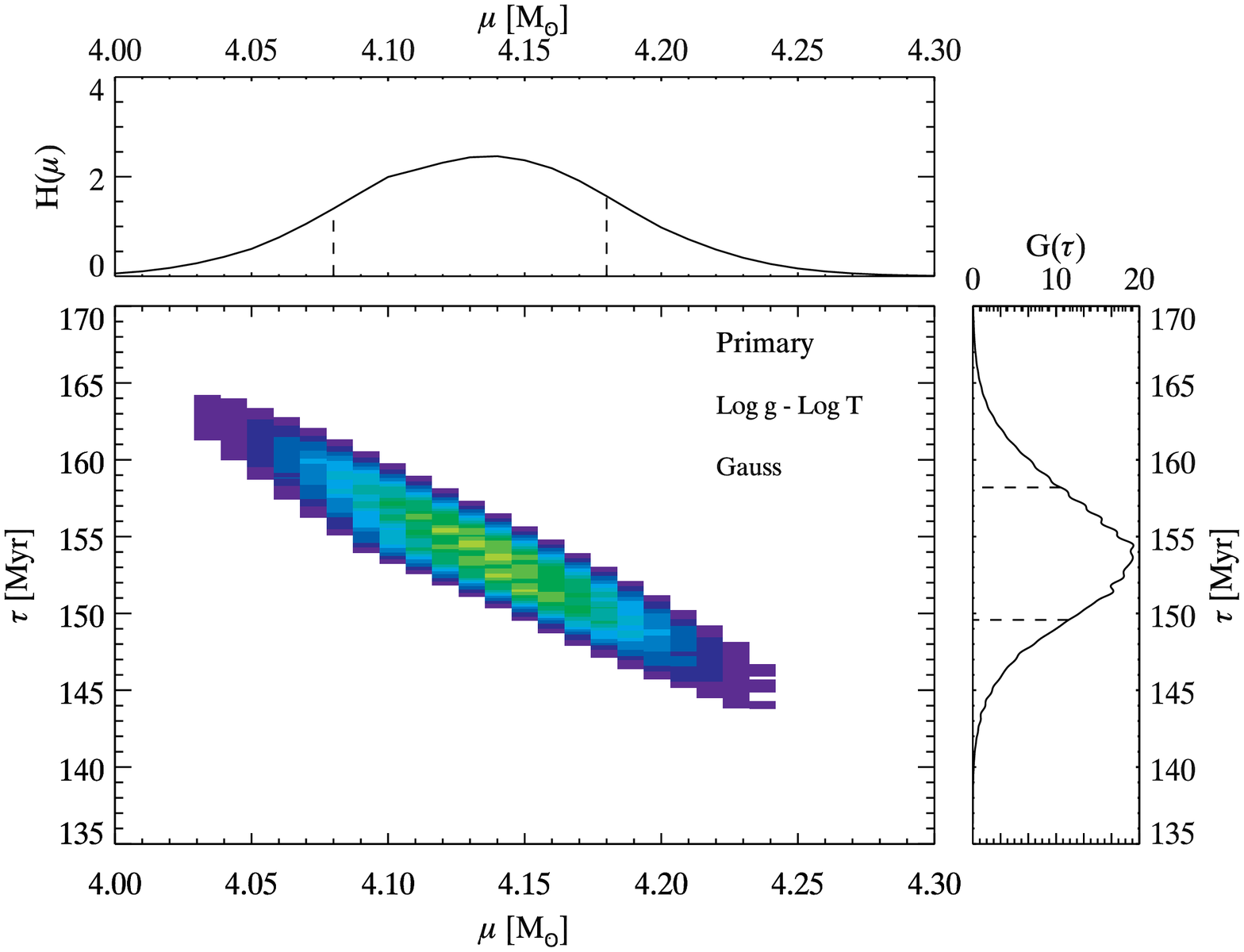} \\  
\label{plotP}
\end{array}$
\end{center}
\caption{Posterior probability distributions for the primary component according to the set 
of evolutionary models constructed by assuming $\beta=0.2, \eta=0.4$ in the L--T (left), in 
the M--R (middle) and in the g--T (right) plane using either a flat (top panels) or a 
Gaussian (bottom panels) prior on the mass distribution. Each 2D probability is obtained after 
marginalizing over the metallicity Gaussian distribution for LMC Cepheids.
In each panel the probability was normalized to its maximum value for displaying purposes,  
therefore the color coding range from 0 (blue) to 1 (red) normalized probability.
The top and the right insets display the marginalization for the stellar age and the stellar 
mass of the primary, respectively. The vertical dashed lines show the confidence intervals. 
The y-axis of the insets show the actual value of the marginal distributions before the 
normalization.}     
\end{figure*}

\begin{figure*}[!h]
\begin{center}$
\begin{array}{ccc}
\includegraphics[width=0.32\textwidth]{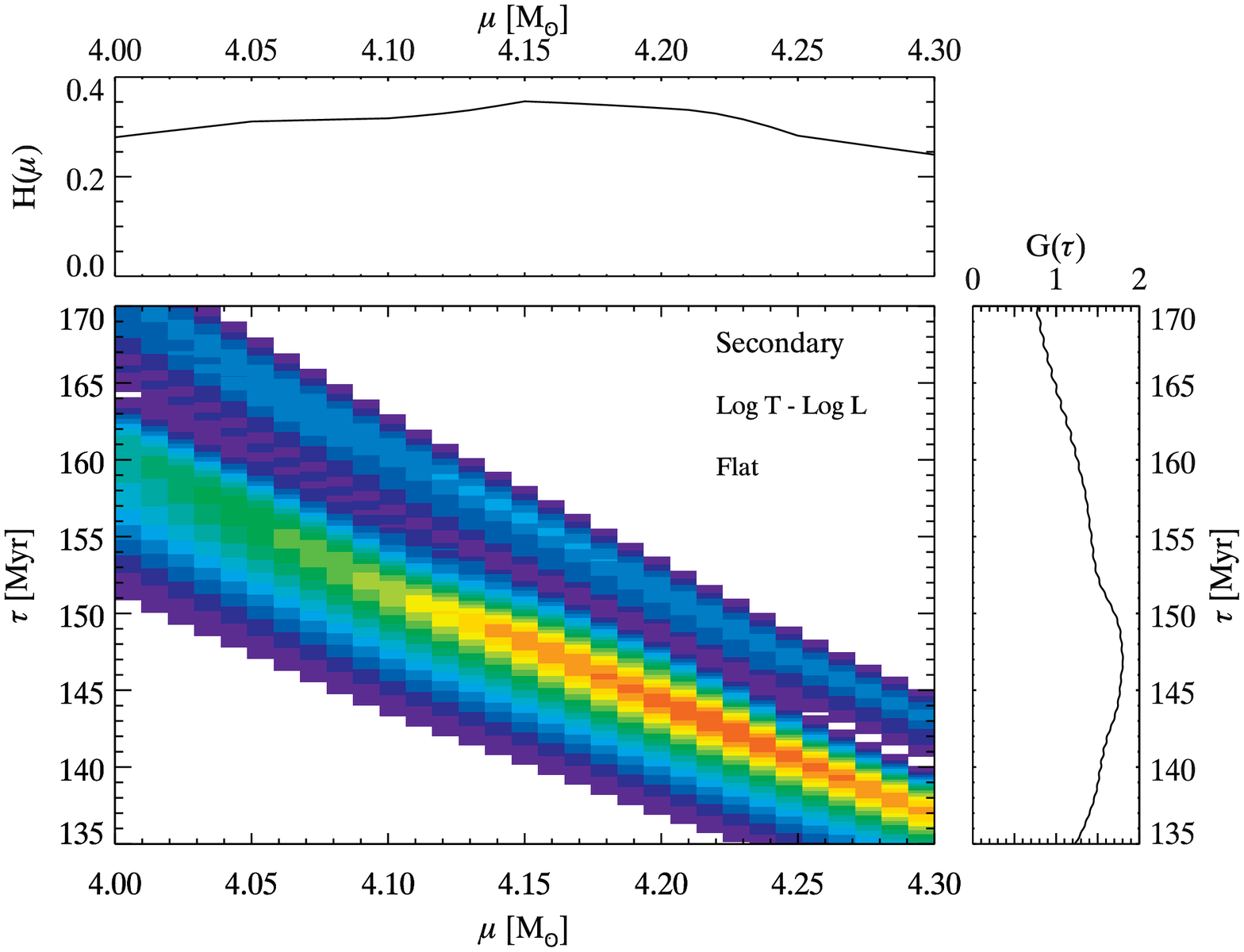} & 
\includegraphics[width=0.32\textwidth]{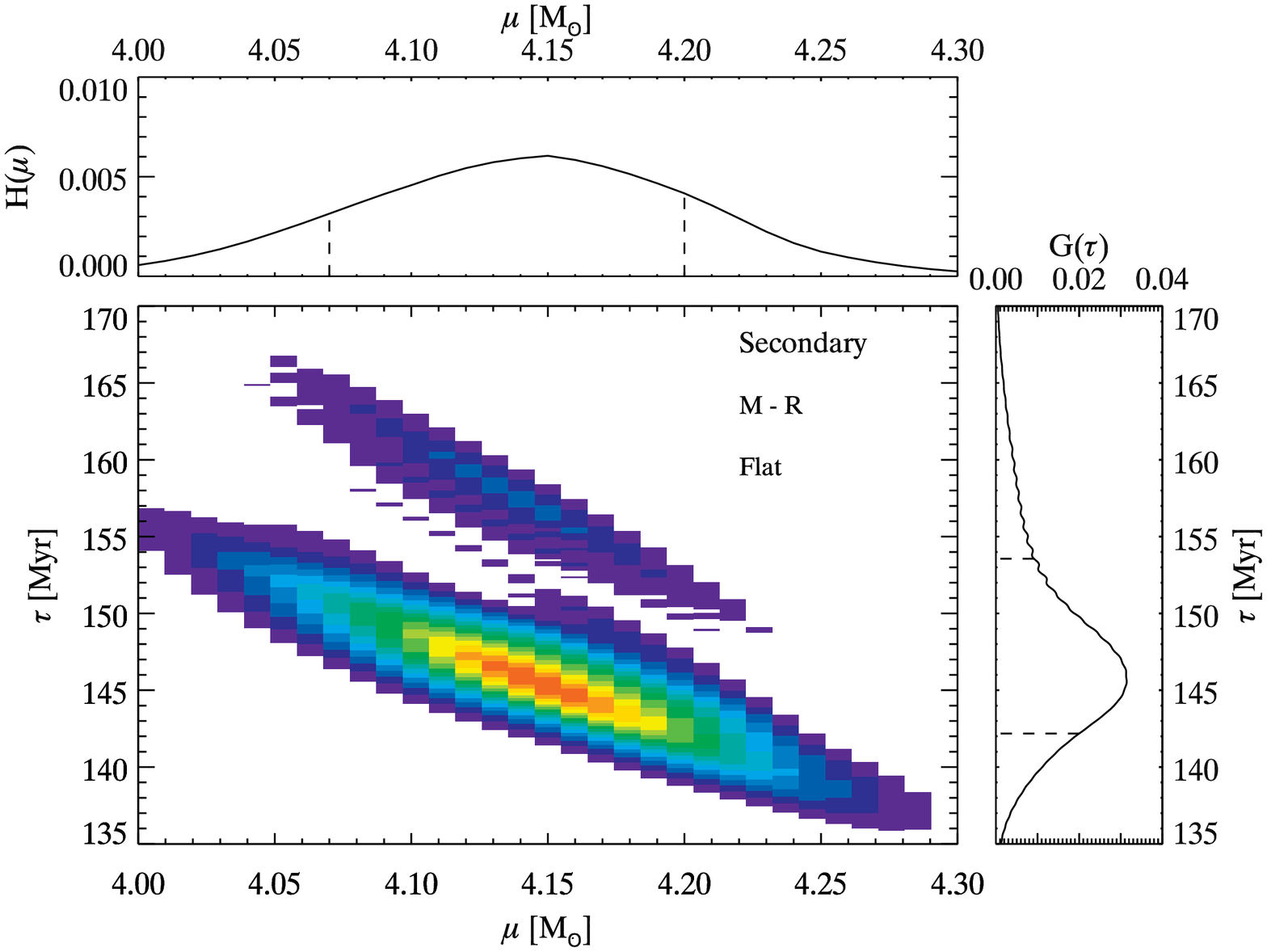} & 
\includegraphics[width=0.32\textwidth]{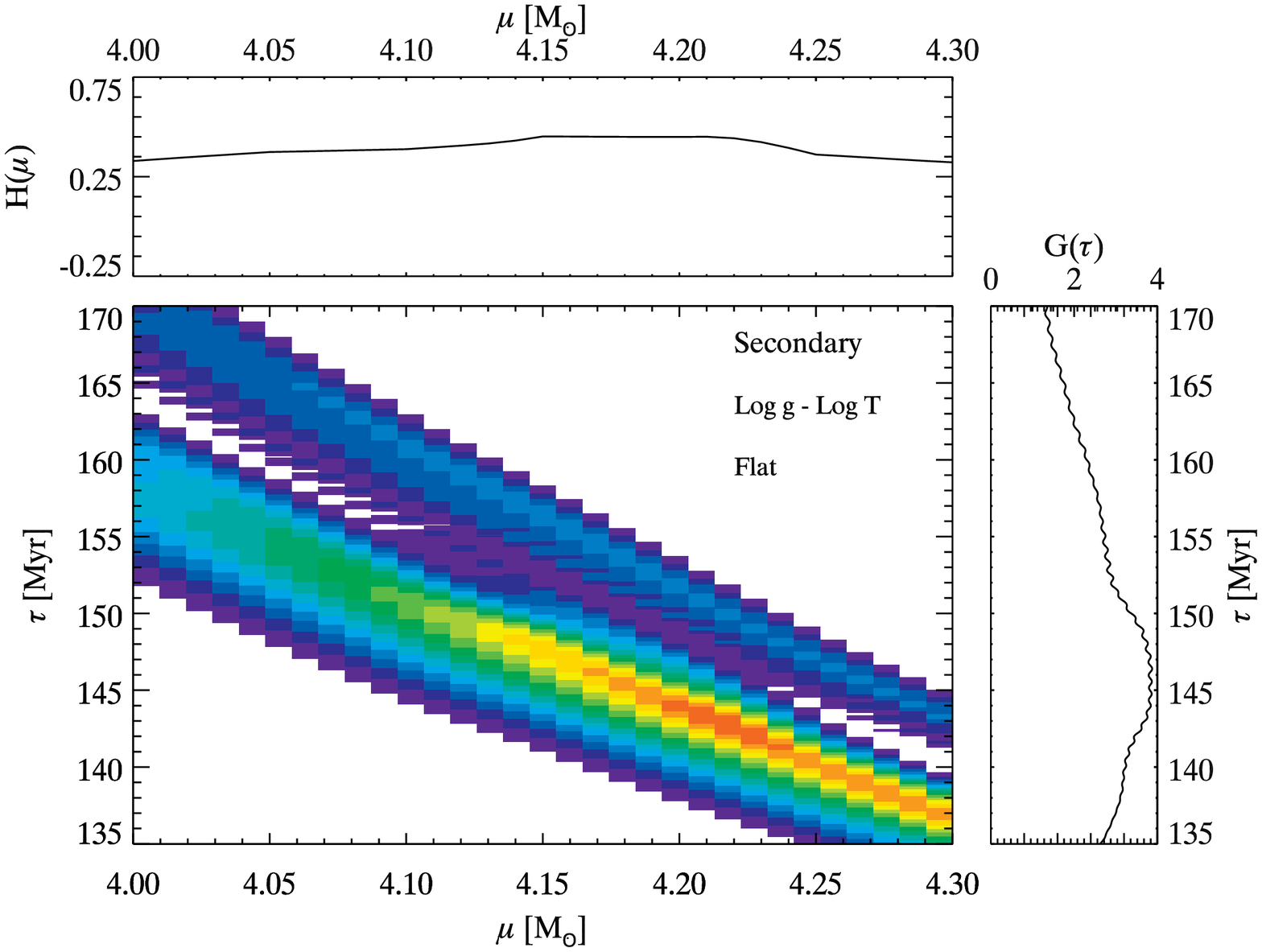} \\  
\includegraphics[width=0.32\textwidth]{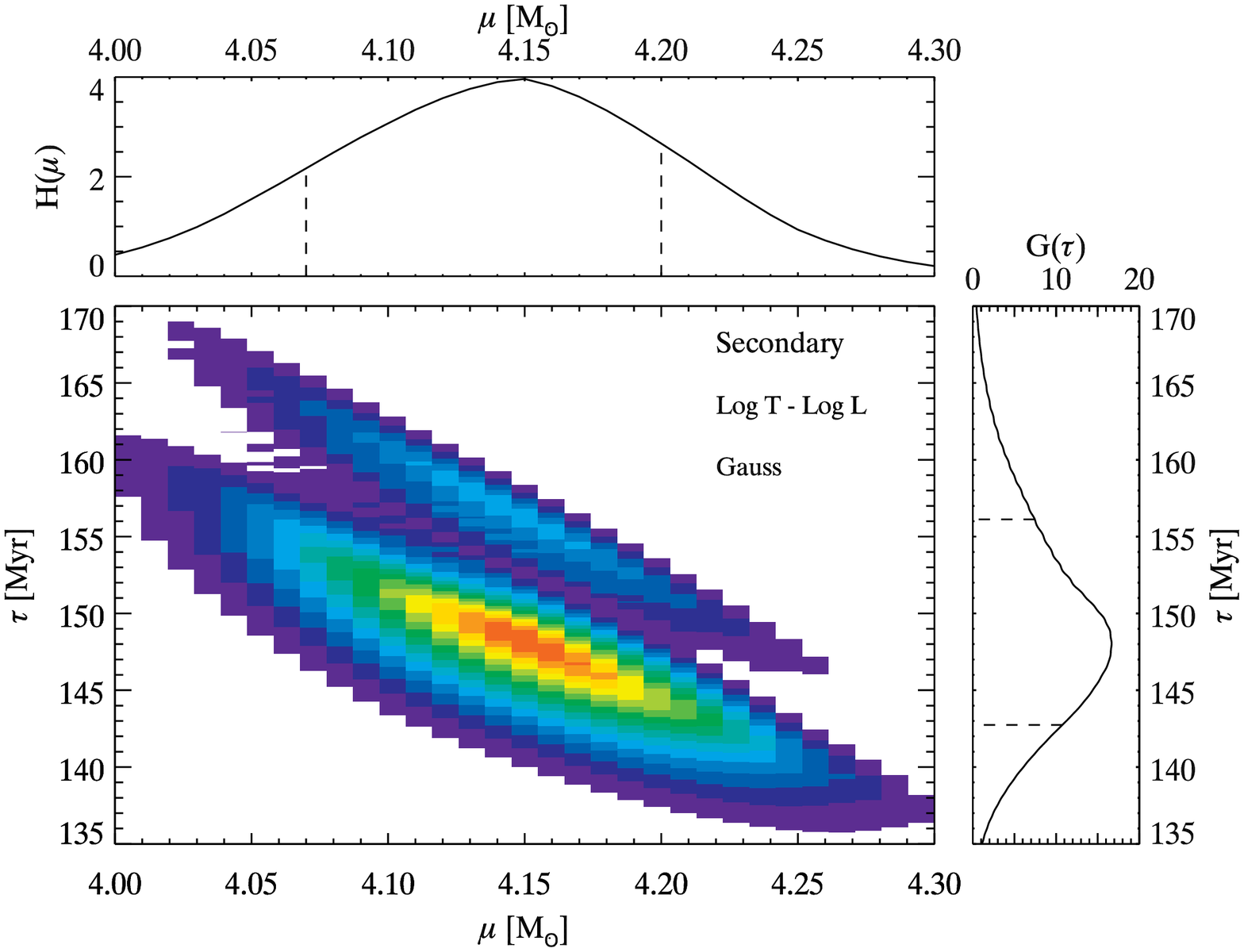} & 
\includegraphics[width=0.32\textwidth]{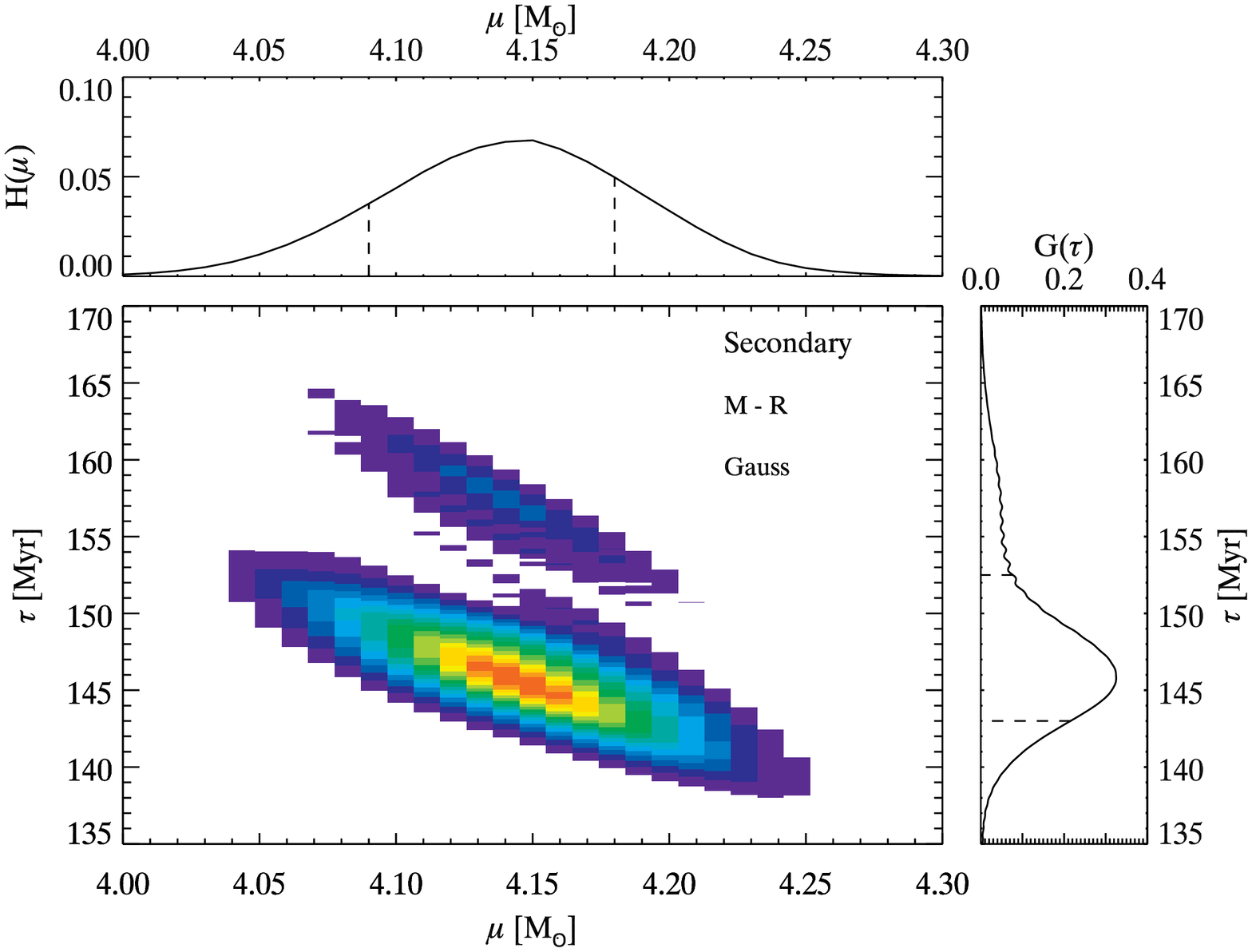} & 
\includegraphics[width=0.32\textwidth]{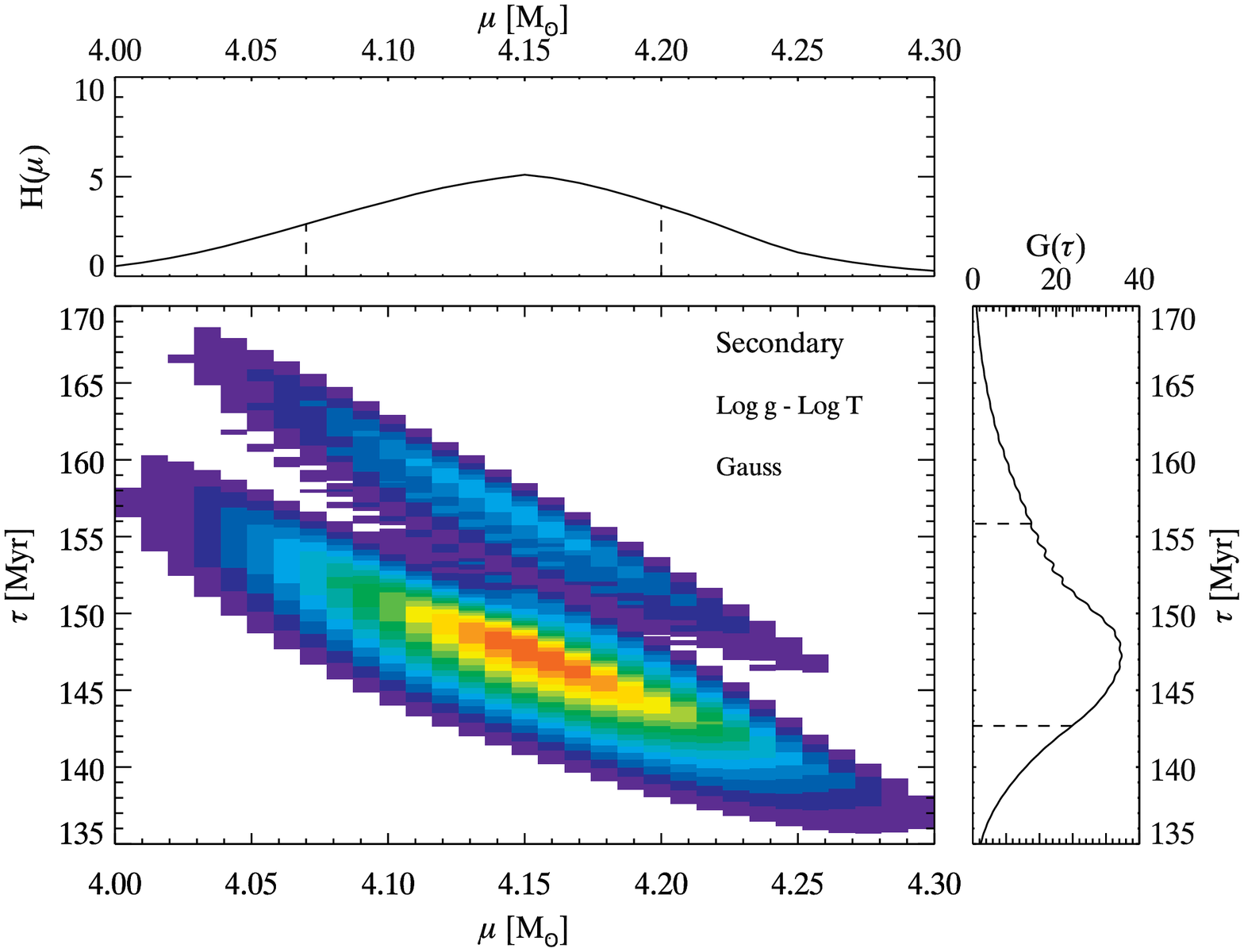}   
\label{plotS}
\end{array}$
\end{center}
\caption{Same as Fig.~6, but for the secondary component.} 
\end{figure*}

 
Finally, we mention that the presence of two different regions in 
all the probability diagrams for the secondary component (see Fig.~7) is the 
consequence of an intrinsic evolutionary feature during the blue loop 
(helium burning phases). Intermediate-mass stars soon after helium ignition 
(tip of the red giant branch)  move, at fixed stellar mass, towards higher 
effective temperatures (the so-called blue tip) and then back to lower 
effective temperatures \citep[][and references therein]{bon00}. 
During these phases stellar structures with similar 
masses can be found in similar positions at different ages. However, the
contribution of the two probability-regions to the marginal distribution
is very different, with the main region contributing a much larger 
probability than the secondary one. The latter region only causes an 
increase in the skewness of the marginal distributions, without 
manifestation of secondary maxima.

\section{The color-magnitude diagram}

\begin{figure}[!t] 
\includegraphics[width=.99\columnwidth]{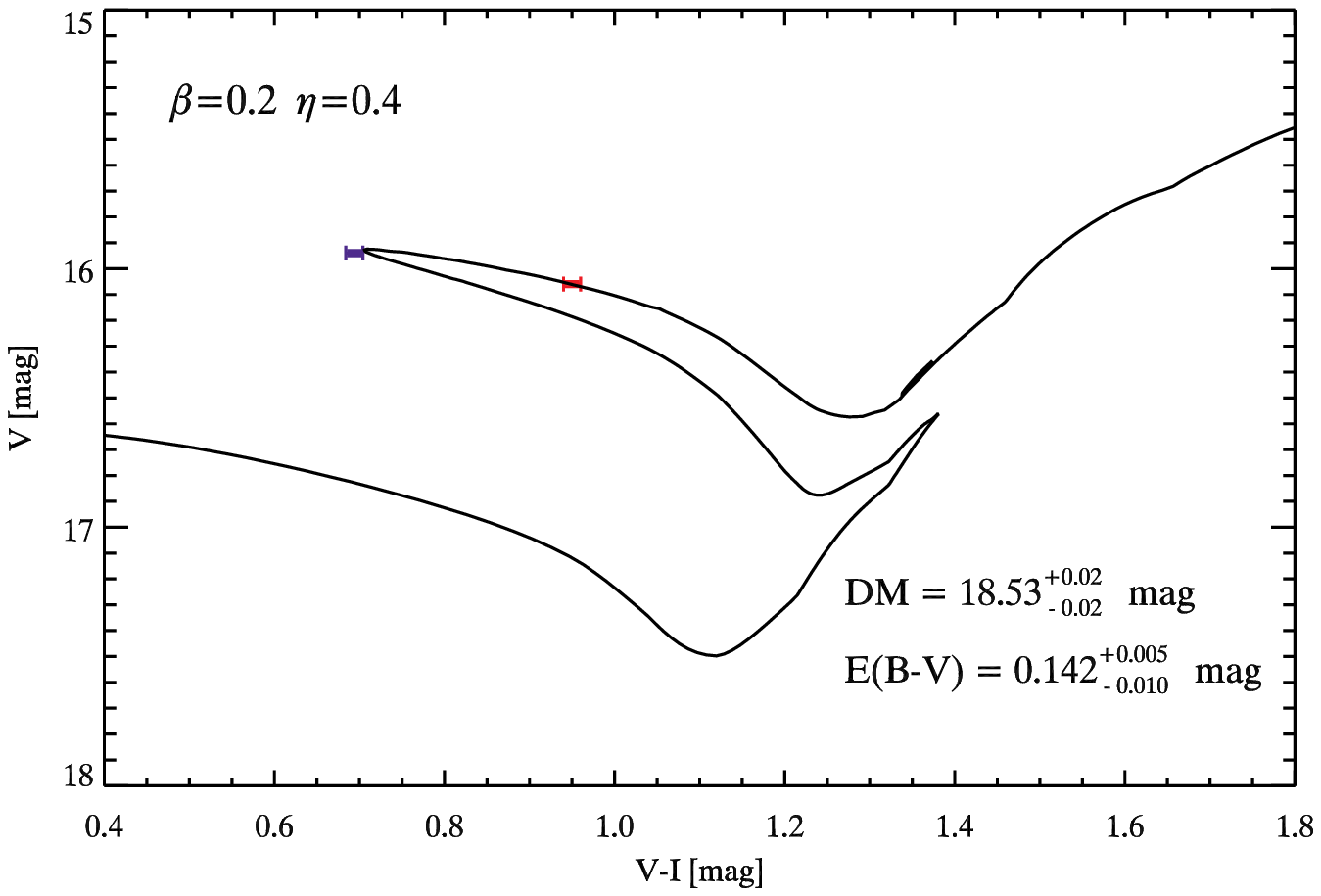}
\label{plotP}
\caption{The isochrone of the most probable solution ($\beta$=0.2, $\eta$=0.4,
g--T plane, Gaussian mass prior) transformed into the V,V-I color-magnitude diagram. 
The blue and the red symbols display the Primary (Cepheid) and the Secondary 
component of the binary system. The true distance modulus and the reddening estimated 
using the combined system are also labeled.} 
\end{figure}

The binary system CEP0227 offers a unique opportunity to test the plausibility 
of the physical assumptions adopted in evolutionary and pulsation models because 
the comparison between theory and observations can be performed by using 
observables (luminosity, radius, effective temperature, surface gravity) 
directly predicted by evolutionary models. This means that these observables 
do not need to be transformed into the observational plane. However, the 
Cepheid mass discrepancy problem arises from the comparison between 
evolutionary prescriptions and observations in the CMD.        
The main advantage in using this plane to estimate the evolutionary mass 
is that the estimate only relies on two mean magnitudes in different 
photometric bands. The main drawback is that the mass estimates are 
affected by uncertainties in the true distance modulus and in the reddening. 
To overcome the latter problem the mass estimates have also been performed 
in CMDs based on NIR bands \citep{bon01}. 

Absolute distance determination and NIR magnitudes for the binary 
CEP0227 are not available yet. We have already collected new NIR time 
series data and we plan to provide a detailed analysis of the CMD 
and of the color-color plane in a forthcoming investigations.

However, the solutions based on theoretical observables do 
allow us to constrain the possible occurrence of thorny systematic 
uncertainties in the bolometric corrections and in the color-temperature 
relations adopted to transform theoretical 
predictions into the observational plane. Moreover, we can also 
estimate the reddening of the system by assuming as a prior 
a Gaussian distribution on the true distance modulus.  
We used the theoretical solution which gives the largest $\mathrm{evidence}$   
for the composite system, i.e. the 151 Myr isochrone for the 
$\beta=0.2, \eta=0.4$ set. This is the most probable age for 
the composite $G(\tau)$ in the case of the g--T plane.
We transformed the isochrone into the Johnson--Kron--Cousins  
photometric system using the spectra provided by \citet{ck03} based on the 
ATLAS9 model atmospheres (see Fig.~8). We also performed a test 
by using the spectra based on the 
atmosphere models provided by the PHOENIX code in the version by \citet{bro05}. 
We found that the difference --in the typical temperature, metallicity and surface 
gravity regime of the binary we are dealing with-- is negligible, therefore in the 
following we adopt the evolutionary models transformed using the ATLAS9 models. 

The comparison into the observational plane (CMD) between the isochrone of the most probable 
solution  and observations provides independent estimates of both the reddening and the 
true distance modulus of the binary system. The reddening in the V,I-bands was 
fixed by adopting the \citet{car89} extinction law.  The solution was found 
by maximizing the probability:

\begin{eqnarray}
\label{eq:fitdmebv}
  p(E(B-V),DM ) =  \prod_{i = P,S} \frac{1}{2\pi\sigma_{V_i}\sigma_{(V-I)_i}}  \times \\ \nonumber
  \int \exp \left[ -\frac{\chi_{V_i}^2 + \chi_{(V-I)_i}^2 }{2} - \frac{(DM-DM_0)^2}{2\sigma^2_{DM_0}} \right] \mathrm{d}m
\end{eqnarray}
where:
\[
\chi_{V,i} = \frac{ V_i - V(m; E(B-V), DM ) }{ \sigma_{V_i} } 
\]
and
\[
\chi_{(V-I)_i} = \frac{ (V-I)_i - (V-I)(m; E(B-V), DM ) }{ \sigma_{(V-I)_i} } 
\]
The index $i$ runs on the primary and on the secondary star $(P,S)$, $V_i$ and $(V-I)_i$ 
are the observed mean magnitudes and colors with their errors $\sigma_{V_i}$ and 
$\sigma_{(V-I)_i}$. The theoretical quantities are function of the mass along the isochrone 
and of the adopted distance modulus and reddening.
The integral is performed along the isochrone integrating over the mass of the models.
Given the available information on the LMC distance modulus, we used a Gaussian prior 
with mean and standard deviations given by $\mu = 18.45\pm0.04$ mag \citep{stor11}.

Fig.~8 shows the 151 Myr isochrone in the V, V-I CMD, with a reddening of 0.142$^{+0.005}_{-0.010}$ mag
and a distance of 18.53$^{+0.02}_{-0.02}$ mag as found by maximizing equation (\ref{eq:fitdmebv}).   
This estimate of the reddening agrees very well with the reddening estimates based on 
Schlegel's map, namely 0.15$\pm$0.03 mag \citep{sch98}. The same outcome applies to the LMC distance, 
since it agrees with similar estimates available in the literature \citep[][and references therein]{stor11}. 
This evidence is also supported by the fact that the plane of the LMC is also tilted along the line of 
sight and we are not applying any correction for the position of the binary system \citep{per04}.  
In order to constrain the sensitivity of both distance modulus and reddening
on the prior we also adopted a flat prior between $\mu$=18 and 19 mag. Note that the 
adopted interval is significantly larger than the range of LMC distances indicated in
literature and we found a true distance modulus of $\mu$=$18.57_{-0.01}^{+0.13}$
mag and a reddening of E(B-V) = $0.1275_{-0.07}^{+0.0025}$ mag. The new
estimates have larger confidence intervals, but they agree quite well with the
estimates based on the Gaussian prior. 
The above findings and the robustness of the most probable solution (see Fig.~9) 
provide an independent support to the current set of atmosphere models adopted to 
transform theory into the observational plane \citep{ck03}.  

\begin{figure}[!t] 
\includegraphics[width=.99\columnwidth]{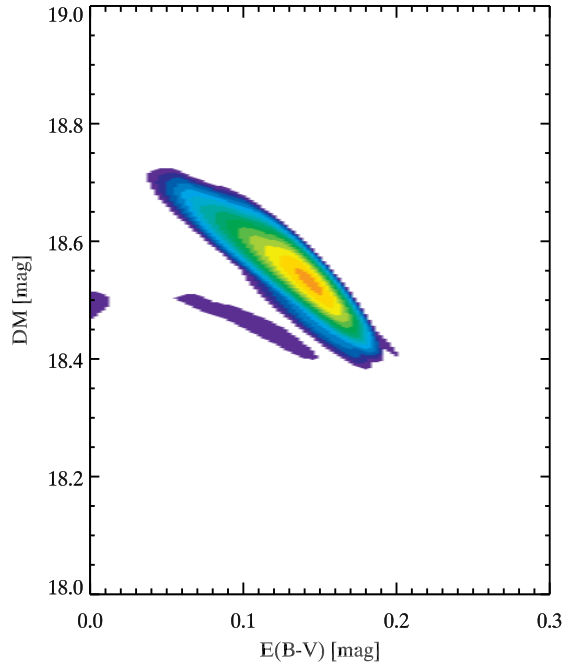}
\label{plotlast}
\caption{Robustness of the most probable solutions for the true distance modulus and the reddening 
of the binary system obtained using the isochrone with the largest $\mathrm{evidence}$ and a prior 
on the true distance modulus (18.45$\pm$0.04) mag. The color coding shows the transition from 
less (blue) to more (red) accurate most probable solutions.} 
\end{figure}

\section{Summary and final remarks}

The discovery by OGLEIII of eclipsing binary systems in the Large Magellanic Cloud hosting 
classical Cepheids provided the unique opportunity of measure their masses with a precision 
ranging from 1\% (CEP0227) to 1.5\% (CEP1812, paper II). It goes without saying that the new 
measurements will allow us to constrain the precision and the plausibility of the the input 
physics and physical assumptions adopted in constructing both evolutionary and pulsation 
models. In this investigation, we addressed the evolutionary mass and the age of the binary 
system CEP0227. We adapted the general Bayesian approach developed by \citet{gen12} to 
constrain the physical parameters of the binary components. 

In order to provide robust statistical solutions, we computed several sets of evolutionary 
models by covering a broad range of chemical compositions (see Table~1) and stellar masses.    
To constrain the dependence on the free parameters currently adopted in canonical evolutionary 
models, independent sets of models were constructed by assuming two different values of the 
mixing length parameter ($\alpha$=1.74, 1.90). Moreover, to account for the possible occurrence of extra-mixing 
during central hydrogen burning phases, the models were constructed either by neglecting or 
by including a moderate convective core overshooting ($\beta_{ov}$=0.2). Finally, to account 
for the mass loss during evolved evolutionary phases the models were constructed either by neglecting 
or by including mass loss. To mimic different efficiencies in the mass loss rate, we constructed 
sets of models accounting either for canonical ($\eta$=0.4,0.8) or for enhanced ($\eta$=4)
mass loss rates (see Table~1).   

To overcome possible deceptive errors in the estimate of the physical parameters, 
the comparison between theory and observations was performed in three different 
planes: luminosity vs effective temperature ($\log$L, $\log$$T_{eff}$), 
mass vs radius (M,R) and surface gravity vs effective temperature 
($\log$g, $\log$$T_{eff}$). We computed an independent solution (mass, age, 
chemical composition) in the three quoted planes for every set of evolutionary 
models we have constructed. Note that individual solutions were computed by 
adopting two different priors for the stellar mass. We adopted either a flat 
distribution over the entire range of stellar masses covered by computations    
or a Gaussian distribution anchored to the mean and the standard deviation of 
the dynamical mass measurements (paper I). We also adopted a Gaussian prior for 
the metallicity, in particular we adopted the mean and the standard deviation 
for LMC Cepheids recently provided by \citet{roma08} by using high-resolution, 
high signal to noise spectra.

The main results we obtained in the comparison between theory and observations are 
the following: 

\begin{itemize}

\item The most probable solutions are minimally affected by the adopted mixing length parameter, 
within the range covered by our models.  

\item The most probable solutions based on a Gaussian mass prior are typically more accurate than 
those based on the flat mass prior.   

\item The most stringent test to constrain the models were obtained in the g--T plane, 
due to the fact that the adopted 
parameters involve three independent observables, namely mass, radius and temperature. In this 
plane, the most probable solutions were obtained by using the evolutionary models that account for 
a moderate convective overshooting ($\beta_{ov}$=0.2) and canonical mass loss rate ($\eta$=0.4). 
In particular, we found that the mass of the primary (Cepheid) is M=4.14$^{+0.04}_{-0.05}$ M$_\odot$, 
while for the secondary is M=4.15$^{+0.04}_{-0.05}$ M$_\odot$. Note that current estimates do agree 
at the 1\% level with dynamical measurements. We found ages for the individual components and for 
the combined system (t=151$^{+4}_{3}$ Myr) do agree at the 5\% level, supporting once again the 
robustness of the solution. It is noteworthy that similar solutions are obtained by using 
moderate convective overshooting ($\beta_{ov}$=0.2) and no mass loss ($\eta$=0). 
This indicates that either a vanishing or a canonical mass loss rate play a marginal 
role in the Cepheid mass discrepancy. Current findings support the preliminary 
results concerning the evolutionary mass of CEP0227 provided by \citet{cas11}.   
They also suggest that evolutionary masses based on evolutionary models 
accounting for moderate convective core overshooting --a typical assumption
in the comparison between theory and observations \citep{pie06}-- agree 
quite well with the dynamical mass of Cep0227. 
The overall agreement among dynamical, evolutionary and pulsation masses
will be addressed in a forthcoming paper in which we plan to constrain the 
pulsation mass using independent observables.

\item The most probable solutions based on the models constructed by assuming a moderate convective 
overshooting ($\beta_{ov}$=0.2) and an enhanced mass loss rate ($\eta$=4) present large 
confidence interval when compared with the solutions based on canonical mass loss rate 
(see Table~3). Note that current finding should be cautiously treated. The reason is twofold. 
The difference between the estimated masses and the dynamical masses, taken at face value, do 
agree at the 1\% level. 
We are assuming a very crude approximation to account for the mass loss, while current 
empirical and theoretical findings indicate that the mass loss might be driven by the 
pulsation instability \citep{nei11,mat11}. 
This hypothesis is further supported by the recent results obtained by       
\citet{neilson2012}. They found that the observed period change of Polaris 
implies a mass loss rate of $\approx$ 10$^{-6}$ M$_\odot$ yr$^{-1}$, and 
in turn a new firm empirical argument for enhanced mass loss in Cepheids. 
\end{itemize} 

The above discussion concerning the formal precision of the results relies on 
the width of the confidence interval. However, a solution characterized by a narrow 
confidence interval might not be the most probable solution. 
The Bayes Factor between two sets of models is a more robust statistical parameter 
to constrain on a quantitative basis the most probable solutions.         

By using the Bayes Factor the main results we obtained in the comparison between theory 
and observations are the following: 

\begin{itemize}

\item The solutions with the largest $\mathrm{evidence}$ were obtained in the g--T plane by using sets of models 
that account for a moderate convective core overshooting ($\beta_{ov}$=0.2) and either 
a vanishing ($\eta$=0) or a canonical mass loss rate ($\eta$=0.4) give very similar 
results. This finding supports the results based on the confidence interval.  

\item The solutions based on sets of models that either neglect convective core 
overshooting or account for an enhanced mass loss rate ($\eta$=4) are characterized by very 
low Bayes Factors. Thus suggesting that these two sets of models give a poorer
description of the observed properties of the binary system. 

\end{itemize} 

We also transformed the isochrone of the most probable solution ($\beta$=0.2, $\eta$=0.4, 
g--T plane, Gaussian mass prior) mentioned above into the observational plane using the 
atmosphere models provided by \citet{ck03}. By assuming a Gaussian metallicity prior 
distribution for the Cepheid metallicity and by assuming the Cardelli  et al. (1989) 
reddening law, we found a true distance modulus --18.53$^{+0.02}_{-0.02}$ mag-- 
and a reddening value --E(B-V)= 0.142$^{+0.005}_{-0.010}$ mag-- that agree quite 
well with similar estimates available in the literature \citep{sch98,per04,stor11}.  
The use of a flat prior on the LMC distance, minimally affects these results. 
A more detailed discussion concerning new independent distances of the binary system 
and of the Cepheid will be addressed in a forthcoming investigation.

Intrinsic variables in eclipsing binary systems are a fundamental laboratory 
to constrain the input physics of both evolutionary and pulsation models. 
This was considered till a few years ago a broken ground for quantitative 
astrophysics, due to the limited number of known systems and the lack of 
homogeneous and precise photometric and spectroscopic data. The unprecedented 
photometric catalogs released by microlensing experiments (EROS, OGLEIII) together 
with currently underway optical (PanSTARRS-1, \citet{stubbs};  Palomar Transient Factory, 
\citet{vane11}; CHASE, \citet{foerster}) and NIR (VVV, \citet{minniti}) surveys 
and the spectrographs available at 8m class telescopes, are a very good viaticum 
in waiting for GAIA.

\acknowledgments
It is a pleasure to thank F. Caputo for many useful discussions concerning 
evolutionary and pulsation properties of classical Cepheids and 
M. Dell'Omodarme for his invaluable help in computer programming. 
We also acknowledge an anonymous referee for his/her positive opinion concerning 
the content of the paper and for his/her sharp and pertinent comments that helped 
us to improve the content and the readability of the manuscript. We gratefully 
acknowledge financial support for this work from the 
Chilean Center for Astrophysics FONDAP 15010003 and the BASAL Centro de 
Astrofisica y Tecnologias Afines (CATA) PFB-06/2007.
Support from the Polish grants N203 387337, the IDEAS Plus, the FOCUS and 
the TEAM subsidies of the Fundation for Polish Science (FNP) are also acknowledged.
During the preparation of this paper M. Gennaro has been a member and would
like to acknowledge the "International Max Planck Research School for Astronomy 
and Cosmic Physics" at the University of Heidelberg, IMPRS-HD, Germany.

{}

%
\begin{deluxetable}{llllllll}
\tabletypesize{\tiny}
\tablewidth{0pt}
\tablecaption{Intrinsic stellar parameters adopted to compute evolutionary tracks.}
\tablehead{
\multicolumn{1}{c}{Z\tablenotemark{a}} &
\multicolumn{1}{c}{$Y$\tablenotemark{a}} &
\multicolumn{1}{c}{$\alpha_{ml}$\tablenotemark{b}} &
\multicolumn{1}{c}{$\beta_{ov}$\tablenotemark{c}} &  
\multicolumn{3}{c}{$\eta$\tablenotemark{d}} 
}
\startdata
0.004& 0.256 & 1.74 & 0.0  & 0 & 0.4 & 0.8 & 4.0\\
0.005& 0.258 & 1.74 & 0.0  & 0 & 0.4 & 0.8 & 4.0\\
0.006& 0.260 & 1.74 & 0.0  & 0 & 0.4 & 0.8 & 4.0\\
0.007& 0.262 & 1.74 & 0.0  & 0 & 0.4 & 0.8 & 4.0\\
0.008& 0.265 & 1.74 & 0.0  & 0 & 0.4 & 0.8 & 4.0\\

0.004& 0.256 & 1.90 & 0.0  & 0 & 0.4 & 0.8 & 4.0\\
0.005& 0.258 & 1.90 & 0.0  & 0 & 0.4 & 0.8 & 4.0\\
0.006& 0.260 & 1.90 & 0.0  & 0 & 0.4 & 0.8 & 4.0\\
0.007& 0.262 & 1.90 & 0.0  & 0 & 0.4 & 0.8 & 4.0\\
0.008& 0.265 & 1.90 & 0.0  & 0 & 0.4 & 0.8 & 4.0\\

0.004& 0.256 & 1.74 & 0.2  & 0 & 0.4 & 0.8 & 4.0\\
0.005& 0.258 & 1.74 & 0.2  & 0 & 0.4 & 0.8 & 4.0\\
0.006& 0.260 & 1.74 & 0.2  & 0 & 0.4 & 0.8 & 4.0\\
0.007& 0.262 & 1.74 & 0.2  & 0 & 0.4 & 0.8 & 4.0\\
0.008& 0.265 & 1.74 & 0.2  & 0 & 0.4 & 0.8 & 4.0\\

0.004& 0.256 & 1.90 & 0.2  & 0 & 0.4 & 0.8 & 4.0\\
0.005& 0.258 & 1.90 & 0.2  & 0 & 0.4 & 0.8 & 4.0\\
0.006& 0.260 & 1.90 & 0.2  & 0 & 0.4 & 0.8 & 4.0\\
0.007& 0.262 & 1.90 & 0.2  & 0 & 0.4 & 0.8 & 4.0\\
0.008& 0.265 & 1.90 & 0.2  & 0 & 0.4 & 0.8 & 4.0\\
\enddata
\tablenotetext{a}{Metal (Z) and helium (Y) abundance by mass.}
\tablenotetext{b}{Mixing length parameter.}
\tablenotetext{c}{Overshooting parameter.} 
\tablenotetext{d}{Mass loss parameter.} 
\end{deluxetable}

\begin{deluxetable}{lll}
\tabletypesize{\tiny}
\tablewidth{0pt}
\tablecaption{Adopted intrinsic parameters for the binary system CEP0227.}
\tablehead{
\multicolumn{1}{c}{Parameter} &
\multicolumn{1}{c}{Value} &
\multicolumn{1}{c}{Ref.\tablenotemark{a}} 
}
\startdata
Mass ($M/M_\odot$)\tablenotemark{b}   & 4.14$\pm$0.05&  1 \\
\ldots                                & 4.14$\pm$0.07&  1 \\

Radius ($R/R_\odot$)\tablenotemark{b} & 32.4$\pm$1.5 &  1  \\
\ldots                                & 44.9$\pm$1.5 &  1  \\

$T_{eff}$ (K)\tablenotemark{b}        & 5,900$\pm$250&  1  \\
\ldots                                & 5,080$\pm$270&  1  \\

$V$(mag)\tablenotemark{b,c}             & 15.940$\pm$0.001&    \\
\ldots                                & 16.06$\pm$0.02&    \\

$I$(mag)\tablenotemark{b,c}             & 15.246$\pm$0.001&    \\
\ldots                                & 15.11$\pm$0.01&    \\

$\mu$ (mag)\tablenotemark{d}         &18.45$\pm$0.04& 2   \\ 

$E(B-V)$(mag)\tablenotemark{e}      & 0.15$\pm$0.03&  3 \\

[Fe/H] (dex)\tablenotemark{f}        &-0.33$\pm$0.05& 4   \\ 

\enddata
\tablenotetext{a}{References: 1) Pietrzynski et al. (2010); 
2) Storm et al. (2011); 3) Schlegel et al. (1998); 4) Romaniello et al. (2008).}
\tablenotetext{b}{The first line, for each parameter, refers to the primary 
(Cepheid), while the second one to the secondary (companion).}
\tablenotetext{c}{Apparent $V$ and $I$-band magnitudes. The mean magnitude 
of the Cepheid was estimated in flux and then transformed into magnitude.} 
\tablenotetext{d}{True distance modulus.} 
\tablenotetext{e}{Reddening according to the Schlegel's map.} 
\tablenotetext{f}{Mean iron abundance (scaled-solar mixture).} 
\end{deluxetable}

\begin{deluxetable}{llccccccccc}
\tabletypesize{\tiny}
\tablewidth{0pt}
\tablecaption{Most probable ages and masses for the different observables, the different sets 
of evolutionary models and for the priors adopted for the stellar mass distribution.}
\tablehead{
\multirow{5}{*}{MP\tablenotemark{a}} & \multirow{5}{*}{Obs.\tablenotemark{b}} & \multirow{5}{*}{Star\tablenotemark{c}} & \multicolumn{2}{c}{$\beta=0.0, \eta =0$} & \multicolumn{2}{c}{$\beta =0.2, \eta = 0$} & \multicolumn{2}{c}{$\beta =0.2, \eta=0.4$}  & \multicolumn{2}{c}{$\beta =0.2, \eta=4.0$} \\
& & & & & & & & & &  \\[-1mm]
& & & Age\tablenotemark{d} & Mass\tablenotemark{d} & Age\tablenotemark{e} & Mass\tablenotemark{e}  & Age\tablenotemark{f} & Mass\tablenotemark{f} & Age\tablenotemark{g} & Mass\tablenotemark{g}\\ 
& & & & & & & & & &  \\[-1mm]
& & & [Myr] & $[M_\odot]$ & [Myr] & $[M_\odot]$  & [Myr] & $[M_\odot]$ & [Myr] & $[M_\odot]$
}
\startdata
 & & & & & & & & & & \\[-1mm]
\multirow{20}{*}{Flat} & \multirow{5}{*}{L--T} & P &$115_{-9}^{+44}$ & $4.50_{-0.52}^{+0.18}$ & $165_{-19}^{+32}$ & $3.98_{-0.29}^{+0.24}$ & $167_{-22}^{+29}$ & $3.97_{-0.27}^{+0.27}$ & $137_{\it -42\tablenotemark{h}}^{+95}$ & $3.60_{-0.15}^{+1.05}$ \\
 & & & & & & & & & & \\[-1mm]
& & S &$96_{-6}^{+20}$ & $4.70_{-0.28}^{+0.16}$ & $143_{-11}^{+36}$ & $4.15_{-0.33}^{+0.19}$ & $147_{-15}^{+30}$ & $4.15_{-0.32}^{+0.20}$ & $148_{-12}^{+26}$ & $4.19_{-0.33}^{+0.14}$ \\
 & & & & & & & & & & \\[-1mm]
& & C &$107_{-10}^{+12}$ &  & $154_{-13}^{+21}$ &  & $150_{-9}^{+24}$ &  & $137_{\it -3\tablenotemark{h}}^{+67}$ &  \\
 & & & & & & & & & & \\[-1mm]
 & & & & & & & & & & \\[-1mm]
 & & & & & & & & & & \\[-1mm]

& \multirow{5}{*}{M--R} & P &$134_{-6}^{+11}$ & $4.13_{-0.05}^{+0.04}$ & $154_{-5}^{+4}$ & $4.10_{-0.02}^{+0.07}$ & $153_{-4}^{+5}$ & $4.14_{-0.05}^{+0.04}$ & $145_{\it -11\tablenotemark{h}}^{+21}$ & $4.14_{-0.07}^{+0.05}$ \\
 & & & & & & & & & & \\[-1mm]
& & S &$114_{-4}^{+30}$ & $4.19_{-0.09}^{+0.05}$ & $145_{-3}^{+7}$ & $4.14_{-0.07}^{+0.05}$ & $146_{-4}^{+7}$ & $4.15_{-0.08}^{+0.04}$ & $146_{-4}^{+7}$ & $4.15_{-0.07}^{+0.05}$ \\
 & & & & & & & & & & \\[-1mm]
& & C &$144_{-7}^{+3}$ &  & $150_{-3}^{+5}$ &  & $150_{-3}^{+5}$ &  & $145_{-0}^{+15}$ &  \\
 & & & & & & & & & & \\[-1mm]
 & & & & & & & & & & \\[-1mm]
 & & & & & & & & & & \\[-1mm]

 & \multirow{5}{*}{g--T} & P &$115_{-14}^{+40}$ & $4.50_{-0.48}^{+0.25}$ & $166_{-17}^{+43}$ & $3.98_{-0.37}^{+0.21}$ & $167_{-21}^{+39}$ & $3.96_{-0.34}^{+0.26}$ & $226_{-22}^{+20}$ & $3.45_{-0.09}^{+0.30}$ \\
 & & & & & & & & & & \\[-1mm]
& & S &$88_{-3}^{+19}$ & $4.98_{-0.44}^{\it +0.0\tablenotemark{h}}$ & $143_{-11}^{+41}$ & $4.15_{-0.42}^{+0.20}$ & $145_{-13}^{+37}$ & $4.15_{-0.40}^{+0.21}$ & $141_{-6}^{+36}$ & $4.20_{-0.42}^{+0.15}$ \\
 & & & & & & & & & & \\[-1mm]
& & C &$95_{-3}^{+19}$ &  & $154_{-14}^{+25}$ &  & $149_{-9}^{+27}$ &  & $139_{\it -52\tablenotemark{h}}^{+96}$ &  \\
 & & & & & & & & & & \\[-1mm]
\cline{2-11}
 & & & & & & & & & & \\[-1mm]

\multirow{20}{*}{Gauss} & \multirow{5}{*}{L--T} & P &$144_{-7}^{+5}$ & $4.13_{-0.05}^{+0.05}$ & $154_{-4}^{+5}$ & $4.12_{-0.04}^{+0.05}$ & $154_{-4}^{+4}$ & $4.13_{-0.05}^{+0.04}$ & $137_{-3}^{+6}$ & $4.15_{-0.05}^{+0.04}$ \\
 & & & & & & & & & & \\[-1mm]
& & S &$118_{-3}^{+32}$ & $4.17_{-0.07}^{+0.06}$ & $147_{-5}^{+9}$ & $4.15_{-0.08}^{+0.04}$ & $148_{-5}^{+8}$ & $4.15_{-0.08}^{+0.04}$ & $150_{-6}^{+7}$ & $4.14_{-0.07}^{+0.05}$ \\
 & & & & & & & & & & \\[-1mm]
& & C &$145_{-5}^{+4}$ &  & $153_{-5}^{+3}$ &  & $151_{-3}^{+5}$ &  & $141_{-3}^{+11}$ &  \\
 & & & & & & & & & & \\[-1mm]
 & & & & & & & & & & \\[-1mm]
 & & & & & & & & & & \\[-1mm]

 & \multirow{5}{*}{M--R} & P &$134_{-6}^{+11}$ & $4.14_{-0.04}^{+0.02}$ & $154_{-4}^{+3}$ & $4.14_{-0.04}^{+0.03}$ & $153_{-3}^{+4}$ & $4.14_{-0.04}^{+0.03}$ & $145_{\it -7\tablenotemark{h}}^{+17}$ & $4.14_{-0.04}^{+0.03}$ \\
 & & & & & & & & & & \\[-1mm]
& & S &$148_{-34}^{\it +2\tablenotemark{h}}$ & $4.17_{-0.07}^{+0.04}$ & $146_{-3}^{+6}$ & $4.14_{-0.04}^{+0.04}$ & $146_{-3}^{+6}$ & $4.15_{-0.05}^{+0.02}$ & $146_{-3}^{+7}$ & $4.14_{-0.04}^{+0.05}$ \\
 & & & & & & & & & & \\[-1mm]
& & C &$146_{-7}^{+2}$ &  & $150_{-2}^{+4}$ &  & $150_{-2}^{+4}$ &  & $145_{-0}^{+13}$ &  \\
 & & & & & & & & & & \\[-1mm]
 & & & & & & & & & & \\[-1mm]
 & & & & & & & & & & \\[-1mm]

 & \multirow{5}{*}{g--T} & P &$145_{-7}^{+5}$ & $4.13_{-0.05}^{+0.05}$ & $154_{-4}^{+5}$ & $4.11_{-0.04}^{+0.05}$ & $154_{-4}^{+4}$ & $4.14_{-0.05}^{+0.04}$ & $137_{-3}^{+19}$ & $4.15_{-0.08}^{+0.04}$ \\
 & & & & & & & & & & \\[-1mm]
& & S &$148_{-32}^{+6}$ & $4.17_{-0.07}^{+0.06}$ & $147_{-5}^{+8}$ & $4.15_{-0.08}^{+0.04}$ & $147_{-4}^{+8}$ & $4.15_{-0.08}^{+0.04}$ & $149_{-5}^{+7}$ & $4.14_{-0.05}^{+0.07}$ \\
 & & & & & & & & & & \\[-1mm]
& & C &$146_{-4}^{+5}$ &  & $153_{-5}^{+3}$ &  & $151_{-3}^{+4}$ &  & $141_{-2}^{+14}$ &  \\
 & & & & & & & & & & \\[-1mm]
\enddata
\tablenotetext{a}{Prior on the stellar mass distribution, either flat or Gaussian distribution.}  
\tablenotetext{b}{Observables adopted in the solution to constrain the age and the mass of the binary system:
L--T: luminosity and temperature; M--R: stellar mass and radius; g--T: surface gravity and temperature.}  
\tablenotetext{c}{Solution for the primary (P, Cepheid), secondary (S, red giant), combined (C).}  
\tablenotetext{d}{Solution for stellar age (Myr) and stellar mass (solar units) based on evolutionary models 
with an overshooting parameter $\beta$=0 (no overshooting) and a mass--loss parameter $\eta=0$ (no mass--loss).}
\tablenotetext{e}{Solution based on evolutionary models with $\beta$=0.2 (mild overshooting) and $\eta$=0 (no mass-loss).}
\tablenotetext{f}{Solution based on evolutionary models with $\beta$=0.2 (mild overshooting) and $\eta$=0.4 (canonical mass-loss rate).}
\tablenotetext{g}{Solution based on evolutionary models with $\beta$=0.2 (mild overshooting) and $\eta$=4.0 (enhanced mass-loss rate).}
\tablenotetext{h}{Errors in italic indicate that the confidence interval is poorly defined.}
\end{deluxetable}


\begin{deluxetable}{llccccc}
\tabletypesize{\tiny}
\tablewidth{0pt}
\tablecaption{Bayes Factors for the different sets of models and for either a flat or a Gaussian 
mass prior.}
\tablehead{
MP\tablenotemark{a} & Obs.\tablenotemark{b} & Star\tablenotemark{c} & $\beta=0.0, \eta =0$\tablenotemark{d} & $\beta =0.2, \eta = 0$\tablenotemark{e} & $\beta =0.2, \eta=0.4$\tablenotemark{f}  & $\beta =0.2, \eta=4.0$\tablenotemark{g}
}
\startdata
\multirow{14}{*}{Flat} & \multirow{4}{*}{L--T} & P & 0.395 & 0.393 & 0.369 & 0.007 \\    
&  \\[-1mm]                
& & S & 0.367 & 0.247 & 0.255 & 0.248 \\                    
&  \\[-1mm]
& & C & 0.145 & 0.097 & 0.094 & 0.002 \\                    
&  \\[-1mm]
& \multirow{4}{*}{M--R} & P & 0.014 & 0.001 & 0.001 & $< 10^{-3}$ \\
&  \\[-1mm]
& & S & $< 10^{-3}$ & 0.001 & 0.001 & 0.001 \\
&  \\[-1mm]
& & C & $< 10^{-3}$ & $< 10^{-3}$ & $< 10^{-3}$ & $< 10^{-3}$ \\
&  \\[-1mm]
& \multirow{4}{*}{g--T} & P & 0.499 & 0.567 & 0.534 & 0.011 \\
&  \\[-1mm]
& & S & 0.236 & 0.311 & 0.321 & 0.304 \\
&  \\[-1mm]
& & C & 0.117 & 0.177 & 0.171 & 0.003 \\
 &  \\[-1mm]
 &  \\[-1mm]
\cline{2-7}
 &  \\[-1mm]
 &  \\[-1mm]
\multirow{14}{*}{Gauss} & \multirow{4}{*}{L--T} & P & 0.654 & 0.857 & 0.825 & 0.002 \\
&  \\[-1mm]
& & S & 0.128 & 0.758 & 0.800 & 0.844 \\
&  \\[-1mm]
& & C & 0.083 & 0.650 & 0.660 & 0.001 \\
&  \\[-1mm]
& \multirow{4}{*}{M--R} & P & 0.159 & 0.010 & 0.010 & $< 10^{-3}$ \\
&  \\[-1mm]
& & S &  $< 10^{-3}$ & 0.009 & 0.010 & 0.012 \\
&  \\[-1mm]
& & C & $< 10^{-3}$ & $< 10^{-3}$ & $< 10^{-3}$ & $< 10^{-3}$ \\
&  \\[-1mm]
& \multirow{4}{*}{g--T} & P & 0.704 & 1.017 & 1.000 & $< 10^{-3}$ \\
&  \\[-1mm]
& & S & 0.042 & 0.949 & 1.000 & 0.978 \\
&  \\[-1mm]
& & C & 0.030 & 0.965 & 1.000 & $< 10^{-3}$ \\
\enddata
\tablenotetext{a}{Prior on the stellar mass distribution, either flat or Gaussian distribution.}  
\tablenotetext{b}{Observables adopted in the solution to constrain the age and the mass of the binary system:
L--T: luminosity and temperature; M--R: stellar mass and radius; g--T: surface gravity and temperature.}  
\tablenotetext{c}{Solution for the primary (P, Cepheid), secondary (S, red giant), combined (C).}  
\tablenotetext{d}{Solution for stellar age (Myr) and stellar mass (solar units) based on evolutionary models 
with an overshooting parameter $\beta$=0 (no overshooting) and a mass--loss parameter $\eta=0$ (no mass--loss).}
\tablenotetext{e}{Solution based on evolutionary models with $\beta$=0.2 (mild overshooting) and $\eta$=0 (no mass-loss).}
\tablenotetext{f}{Solution based on evolutionary models with $\beta$=0.2 (mild overshooting) and $\eta$=0.4 (canonical mass-loss rate).}
\tablenotetext{g}{Solution based on evolutionary models with $\beta$=0.2 (mild overshooting) and $\eta$=4.0 (enhanced mass-loss rate).}
\end{deluxetable}

\end{document}